\documentclass[journal]{IEEEtran}
\IEEEoverridecommandlockouts
\usepackage{lettrine}
\usepackage{cite}
\usepackage{amsmath,amssymb,amsfonts}
\usepackage{algorithm}
\usepackage{algorithmic}
\usepackage{bm}
\usepackage{graphicx}
\usepackage{subfigure}
\usepackage{tabularx}
\usepackage{textcomp}
\usepackage{xcolor}
\usepackage{stfloats}
\usepackage{multirow}
\def\BibTeX{{\rm B\kern-.05em{\sc i\kern-.025em b}\kern-.08em
		T\kern-.1667em\lower.7ex\hbox{E}\kern-.125emX}}

\floatname{algorithm}{\bf{Algorithm}}

\begin{document}
	
	\title{Intelligent Reflecting Surface Aided Target Localization With Unknown Transceiver-IRS Channel State Information}
	
	\author{Taotao~Ji, Meng~Hua, Xuanhong~Yan, Chunguo~Li,~\IEEEmembership{Senior~Member,~IEEE,} Yongming~Huang,~\IEEEmembership{Senior~Member,~IEEE,} and~Luxi~Yang,~\IEEEmembership{Senior~Member,~IEEE}%
		\IEEEcompsocitemizethanks{\IEEEcompsocthanksitem Taotao Ji, Xuanhong Yan, Chunguo Li, Yongming Huang, and Luxi Yang are with the School of Information Science and Engineering, the National Mobile Communications Research Laboratory, and the Frontiers Science Center for Mobile Information Communication and Security, Southeast University, Nanjing 210096, China, and also with the Pervasive Communications Center, Purple Mountain Laboratories, Nanjing 211111, China (e-mail: jitaotao@seu.edu.cn; yanxuanhong@seu.edu.cn; chunguoli@seu.edu.cn; huangym@seu.edu.cn; lxyang@seu.edu.cn).
			\IEEEcompsocthanksitem Meng Hua is with the Department of Electrical and Electronic Engineering, Imperial College London, London SW7 2AZ, UK (e-mail: m.hua@imperial.ac.uk).
	}}
	\maketitle
	\begin{abstract}
		Integrating wireless sensing capabilities into base stations (BSs) has become a widespread trend in the future beyond fifth-generation (B5G)/sixth-generation (6G) wireless networks. In this paper, we investigate intelligent reflecting surface (IRS) enabled wireless localization, in which an IRS is deployed to assist a BS in locating a target in its non-line-of-sight (NLoS) region. In particular, we consider the case where the BS-IRS channel state information (CSI) is unknown. 
		Specifically, we first propose a separate BS-IRS channel estimation scheme in which the BS operates in full-duplex mode (FDM), i.e., a portion of the BS antennas send downlink pilot signals to the IRS, while the remaining BS antennas receive the uplink pilot signals reflected by the IRS. However, we can only obtain an incomplete BS-IRS channel matrix based on our developed iterative coordinate descent-based channel estimation algorithm due to the ``sign ambiguity issue”. Then, we employ the multiple hypotheses testing framework to perform target localization based on the incomplete estimated channel, in which the probability of each hypothesis is updated using Bayesian inference at each cycle. Moreover, we formulate a joint BS transmit waveform and IRS phase shifts optimization problem to improve the target localization performance by maximizing the weighted sum distance between each two hypotheses. However, the objective function is essentially a quartic function of the IRS phase shift vector, thus motivating us to resort to the penalty-based method to tackle this challenge. 
		Simulation results validate the effectiveness of our proposed target localization scheme and show that the scheme's performance can be further improved by finely designing the BS transmit waveform and IRS phase shifts intending to maximize the weighted sum distance between different hypotheses.
	\end{abstract}
	
	
	\section{Introduction}
	\IEEEPARstart{R} {ecently}, the integration of wireless sensing into the future beyond fifth generation (B5G) and sixth generation (6G) wireless systems to support emerging intelligent applications and services such as machine-type communication (MTC), auto-driving, smart city, industrial automation, extended reality (XR), etc., has attracted growing research interests \cite{10012421_Wei, 9828481_Qi, 10061429_Meng,10178069_Liu}. Toward this end, numerous researchers aim to realize the dual functions of wireless communications and radar sensing based on the sharing of cellular base station (BS) infrastructures and signal processing modules related to sensing.
	It not only improves the efficiency of limited resource utilization but also promotes the interweaving and coordination between communication and sensing in location-aware services and applications to further enhance system performance \cite{9127852_Ma,9966507_Hu}. Among various visions and outlooks for B5G/6G networks, there is a consensus that sensing will play a more crucial role than ever before.
	
	Generally, the operation of wireless sensing relies on line-of-sight (LoS) links between the radar and the targets, such that the sensing information related to the angle of arrival (AoA), including the targets' location and velocity, etc., can be extracted from the echo signals (see, e.g., \cite{923295_Dogandzic} and \cite{4359542_Li}). However, in actual obstruction-dense scenarios, e.g., autonomous driving in a city full of buildings, sensing targets are more likely to be located at the non-LoS (NLoS) region of the radar, causing the conventional LoS-link-dependent sensing no longer applicable \cite{10138058_Song}. At the very least, the presence of a predominantly scattered signal environment will inevitably deteriorate the sensing performance. These naturally raise an open issue: how do wireless sensing networks perform target sensing tasks in harsh environments, such as its NLoS regions?
	
	One of the main purposes for which intelligent reflecting surface (IRS) has been introduced into wireless communication systems is to establish virtual LoS links between a BS and the target users located within its unfavorable service area, thus enabling the communication coverage expansion \cite{10373827_Yang,10278759_Kay,9980412_Shi}.
	Moreover, the IRS offers attractive features such as cost-effectiveness and enhanced deployment flexibility thanks to its low profile, lightweight, and conformal geometry \cite{9771079_Mei,10136805_Ji}.
	Inspired by this, exploiting IRS is regarded as a promising solution for target sensing in the NLoS region of radar. 
	Specifically, by carefully deploying IRS such that the LoS IRS-target link exists and finely adjusting the phase shifts of its reflecting elements, namely reaping passive beamforming gains, to combat the severe signal propagation loss, the radar can perform NLoS target sensing based on the echo signals from radar-IRS-target-IRS-radar link.
	
	Several prior works have investigated IRS-aided wireless sensing \cite{10138058_Song,10284917_Hua,10149442_Pang,9725255_Zhang} and IRS-enhanced integrated sensing and communications (ISAC) \cite{10178069_Liu,9733335_Liu,9769997_Liu,9913311_Hua,10143420_Hua}. Specifically, the authors of \cite{10138058_Song} studied IRS-enabled NLoS wireless sensing, where two types of target models, namely, the point and extended targets, were considered. The active and passive beamforming at the access point (AP) and IRS were jointly designed to minimize the Cramér-Rao bound (CRB) of the estimated targets' direction-of-arrival (DoA) error. The authors of \cite{9725255_Zhang} considered IRS-aided multi-target detection based on multiple hypotheses testing techniques, where the introduction of IRS can enlarge the predicted distance between different hypotheses due to its capability to customize channel conditions, thus improving the detection performance. The authors in \cite{9733335_Liu} optimized the long-term performance of the IRS-aided ISAC system based on the deep reinforcement learning (DRL) method. In \cite{9769997_Liu}, the potential of employing IRS in ISAC systems for enhancing both radar sensing and communication functionalities concurrently was demonstrated. However, the works in \cite{10178069_Liu}, \cite{10138058_Song}, \cite{9733335_Liu}, \cite{9769997_Liu}, \cite{9913311_Hua}, and \cite{10143420_Hua} assumed that the separate BS/AP-IRS channel state information (CSI) is known by default, which is quite difficult to recover from the obtained cascaded channel for passive IRS in practice due to the scaling ambiguity issue \cite{9771077_Swindlehurst,9847080_Pan}. While in \cite{10284917_Hua}, \cite{10149442_Pang}, and \cite{9725255_Zhang}, the authors assumed that the channel between the transceiver and IRS can be calculated based on the propagation characteristics of electromagnetic waves in free space. In conclusion, IRS-empowered wireless sensing with unknown transceiver-IRS CSI is still an open problem, which thus motivates this work.
	
	In this paper, we consider an IRS-aided target localization scenario performed by a BS, where the target is located in the NLoS region of the BS. In particular, we assume that the separate BS-IRS CSI is unknown, which further makes the localization task more general as well as more challenging. In general, our main contributions can be summarized as follows:
	\begin{itemize}
		\item We propose a target localization protocol that coordinates the operations of the BS and the IRS. Specifically, the overall target localization procedure consists of two steps, namely, the separate BS-IRS channel estimation step and the target localization step. According to the developed protocol, the BS and IRS coordinate their respective working mode and phase shifts during different time slots to achieve precise target localization collaboratively. 
		\item In the separate BS-IRS channel estimation stage, we set the BS to full-duplex mode (FDM), where a part of the BS antennas transmit downlink pilot symbols to the IRS, and the remaining BS antennas receive uplink pilot symbols reflected by the IRS concurrently. Based on the transmitted and received pilot signals, we propose an iterative coordinate descent-based algorithm to estimate the BS-IRS channel. Due to the ``sign ambiguity issue", we finally obtain an incomplete BS-IRS channel matrix, i.e., the sign of each row vector is undetermined.
		\item In the target localization stage, we employ the multiple hypotheses testing technique to perform target localization based on the incomplete estimated BS-IRS channel matrix. The probability of each hypothesis is updated using Bayesian inference with each cycle of the BS transmitting localization waveforms. When the termination criterion of the target localization is satisfied, we can obtain a complete BS-IRS channel matrix while acquiring the precise location of the target.
		\item To improve the target localization performance, an optimization problem of maximizing the weighted sum distance between each two hypotheses is formulated, where the BS transmit waveforms and the IRS phase shifts are jointly optimized. It is worth noting that to reduce the complexity of hardware operation, we set the IRS phase shifts in the transmit and receive time slots of the BS to be the same, resulting in the objective function of the formulated problem being a quartic function of the IRS phase shift vector, which further complicates the problem. To tackle this challenge, we resort to the penalty-based method, where all the optimization variables are updated with a closed-form solution.
		\item We present numerical results to validate the performance of our proposed designs. It is shown that our developed target localization scheme without prior BS-IRS CSI can accurately locate the target. Moreover, it is demonstrated that the fine design of the BS transmit waveforms and IRS phase shifts to maximize the weighted sum distance between hypotheses can further improve the target localization performance.
	\end{itemize}
	
	The rest of this paper is organized as follows. In Section~II, we introduce the target localization scenario and propose the target localization protocol. The separate BS-IRS channel estimation method is provided in Section III. In Section IV, we present the target localization scheme in detail. The joint BS transmit waveforms and IRS phase shifts optimization problem is formulated and solved in Section V. In Section VI, we provide simulation results to evaluate our proposed designs. Finally, Section VII concludes the paper.
	
	\begin{figure}[t]
		\centerline{\includegraphics[width=0.4\textwidth]{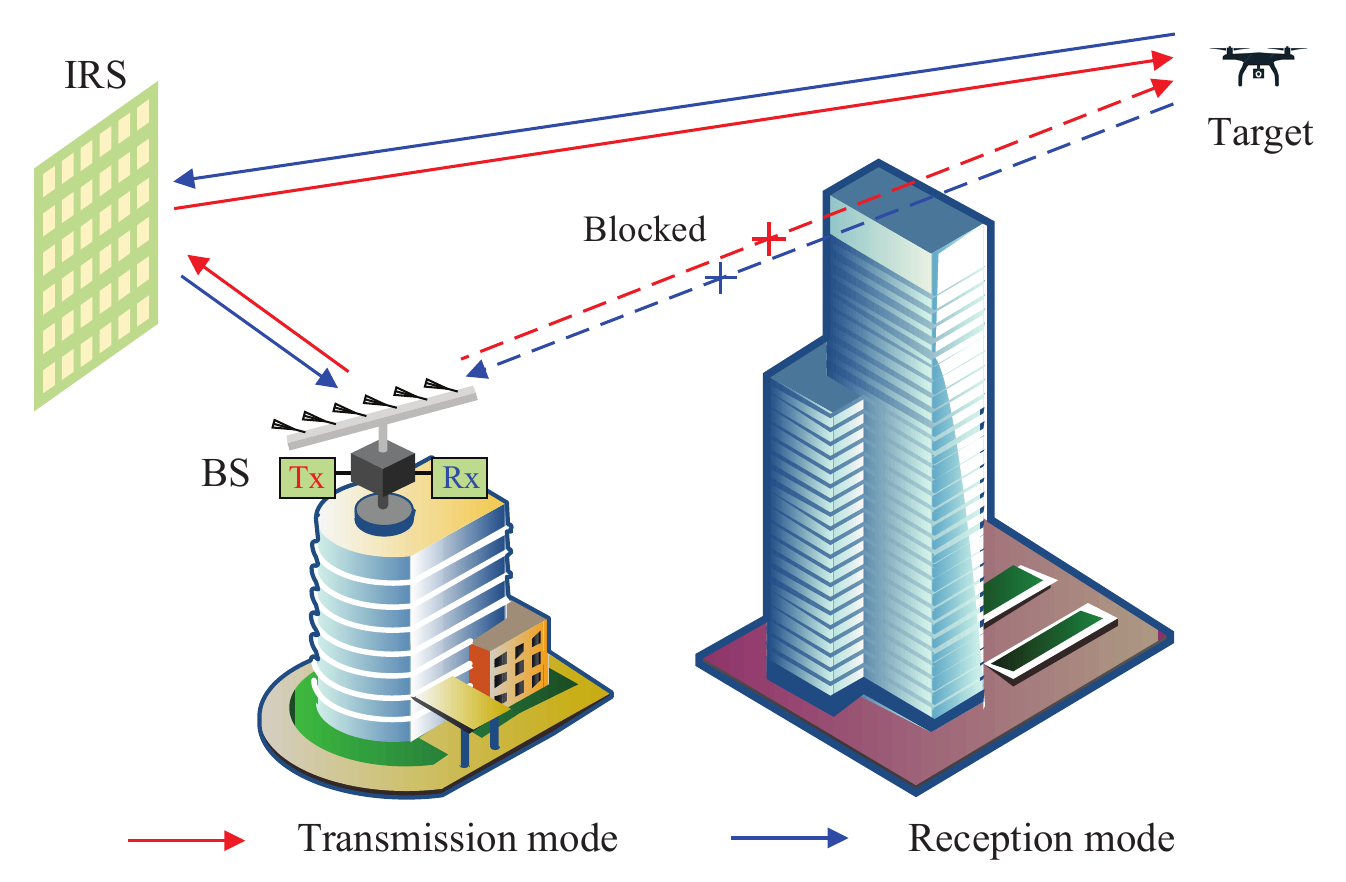}}
		\caption{An IRS-aided target localization system, where the target is in the NLoS region of the BS.}
		\label{fig_system_model}
	\end{figure}
	
	\section{System Model And Target Localization Protocol}
	\subsection{System Model Description}
	As shown in Fig. \ref{fig_system_model}, we consider a scenario in which a BS performs wireless localization of a specific target in the space of interest (SoI) with the assistance of an IRS. In particular, the LoS link between the BS and the target is blocked by obstacles such as buildings. Therefore, by deploying an IRS in the vicinity of the BS, a virtual direct link between the BS and the target can be created. The BS has $ M $ antennas and can operate in three modes, namely, the transmission mode (TM), the reception mode (RM), and the FDM. The IRS is a fully passive device (i.e., there are no sensing devices or radio frequency (RF) chains mounting on it) with $ N $ elements. Denote the set of reflecting elements of the IRS as $ {\cal N} = \left\{ {1,2, \cdots ,N} \right\} $, and the IRS reflection coefficient matrix is modeled as $ {\bf{\Theta }} = {\rm{diag}}\left( {\bm{\theta }} \right) $, where $ {\bm{\theta }} \in {\mathbb{C}^{N \times 1}} $ is the reflection coefficient vector that satisfies $ {\bm{\theta }}\left( n \right) = {e^{j{\phi _n}}},\forall n \in {\cal N} $, with $ {\phi _n} \in \left[ {0,2\pi } \right) $ is the phase shift corresponding to the $ n $-th IRS reflecting element.
	Let $ {\bf{G}} \in {\mathbb{C}^{N \times M}} $ denote the BS-IRS channel, which is assumed to be quasi-static since both the BS and the IRS are placed in fixed positions \cite{9400843_Hu,10136805_Ji}. It should be noted that unlike existing works on IRS-aided wireless sensing, localization, target detection, etc., where the channel between the transceiver and the IRS is precisely known \cite{10178069_Liu,10138058_Song,9733335_Liu,9769997_Liu,9913311_Hua} or can be calculated \cite{10149442_Pang,9725255_Zhang}, whereas in our work, the BS-IRS channel $ \bf{G} $ is unknown beforehand and needs to be estimated for further target localization. Since the spatial extent of the target is small, it can be viewed as a single scatterer. Accordingly, the target response matrix with respect to the IRS is thus modeled as
	\begin{align}\label{TargetRes_matrix}
	{\bf{H}} = \alpha {\bf{a}}\left( {\theta ,\phi } \right){{\bf{a}}^T}\left( {\theta ,\phi } \right),
	\end{align}
	where $ \alpha  \in \mathbb{C} $ represents the complex-valued channel coefficient, which depends on the radar cross section (RCS) of the target and the round-trip path loss of the IRS-target-IRS link, and $ {\bf{a}}\left( {\theta ,\phi } \right) $ denotes the steering vector of the IRS for the target, which is given by
	\begin{align}
	{\bf{a}}\left( {\theta ,\phi } \right) = {{\bf{a}}_x}\left( {\theta ,\phi } \right) \otimes {{\bf{a}}_y}\left( {\theta ,\phi } \right),
	\end{align}
	where
	\begin{subequations}
		\begin{align}
		&{{\bf{a}}_x}\left( {\theta ,\phi } \right) = {\big[ {1,{e^{j\frac{{2\pi {d_x}{\vartheta _x}}}{{{\lambda _c}}}}}, \cdots ,{e^{j\frac{{2\pi {d_x}\left( {{N_x} - 1} \right){\vartheta _x}}}{{{\lambda _c}}}}}} \big]^T},\\
		&{{\bf{a}}_y}\left( {\theta ,\phi } \right) = {\big[ {1,{e^{j\frac{{2\pi {d_y}{\vartheta _y}}}{{{\lambda _c}}}}}, \cdots ,{e^{j\frac{{2\pi {d_y}\left( {{N_y} - 1} \right){\vartheta _y}}}{{{\lambda _c}}}}}} \big]^T},
		\end{align}
	\end{subequations}
	with $ {\vartheta _x} \buildrel \Delta \over = \sin \theta \cos \phi  $, $ {\vartheta _y} \buildrel \Delta \over = \sin \theta \sin \phi  $, $ {d_{x\left( y \right)}} $ and $ {N_{x\left( y \right)}} $ are the inter-element spacing and the number of IRS elements along the $ x\left( y \right) $-axis, respectively, and $ {\lambda _c} $ denotes the carrier wavelength.
	
	\begin{figure}[H]
		\centerline{\includegraphics[width=0.48\textwidth]{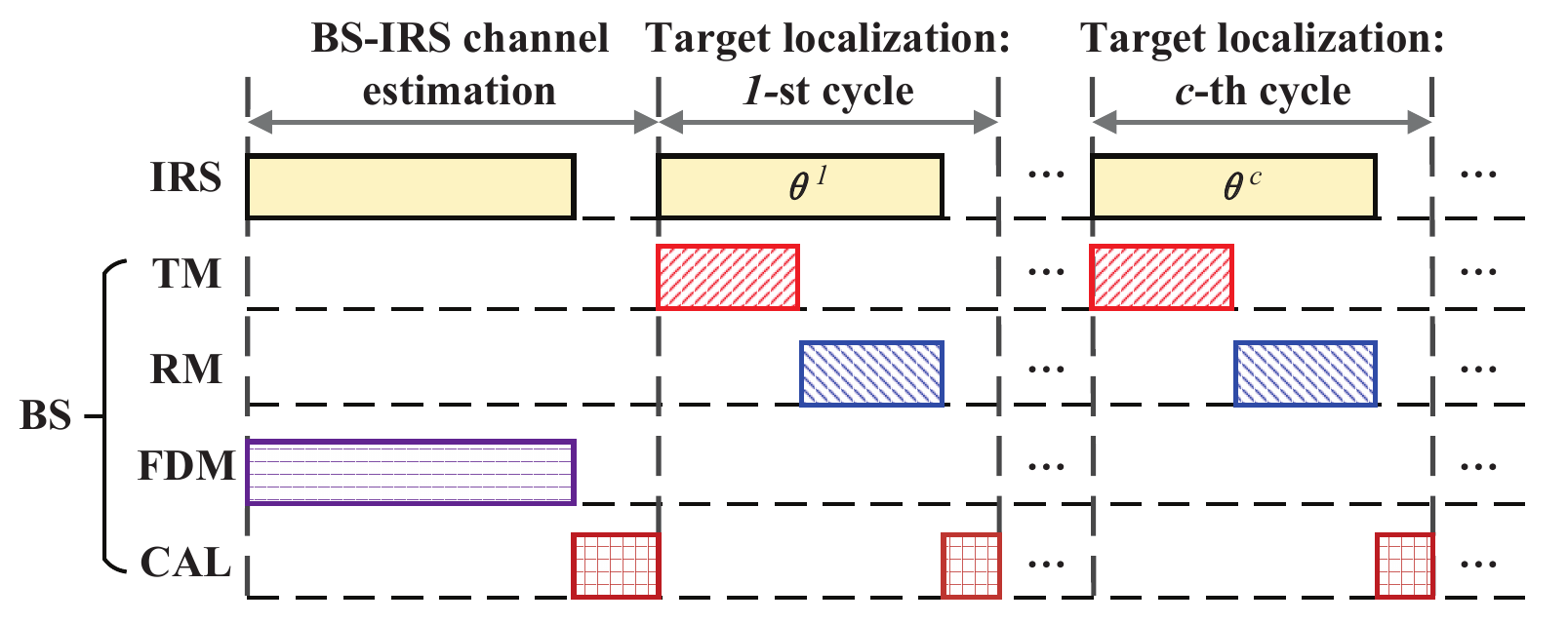}}
		\caption{Proposed target localization protocol.}
		\label{fig_Protocol}
	\end{figure}
	
	\subsection{Proposed Target Localization Protocol}
	In this subsection, we propose a target localization protocol as shown in Fig. \ref{fig_Protocol} to coordinate the operations of the BS and IRS to achieve precise target localization. The whole target localization procedure is divided into two stages, namely, the BS-IRS channel estimation stage and the target localization stage. The details are given as follows.
	
	\subsubsection{BS-IRS Channel Estimation Stage}
	During this stage, the BS is set to FDM, and some of the BS antennas act as transmit antennas to send pilot symbols to the IRS via the downlink channel, while the remaining antennas receive the pilot symbols reflected by the IRS via the uplink channel simultaneously. By setting different transmit/receive antenna combinations, the BS thus obtains a sufficient number of observations. Based on the observation data, the BS estimates the separate BS-IRS channel.
	
	\subsubsection{Target Localization Stage}
	We divide the timeline in the target localization stage into cycles, and each cycle contains three steps, i.e., the transmission, reception, and calculation (CAL) steps. Specifically, we focus on the $ c $-th cycle. In the transmission step, the BS is set to TM and sends the specialized waveforms required for target localization to the IRS. The transmitted signal is then reflected by the IRS to the target. Both the waveforms transmitted by the BS and the IRS phase shifts remain unchanged during this step, which are optimized during the CAL step in the $ \left( {c - 1} \right) $-th cycle. In the reception step, the BS is set to RM and receives the target echo signal via the IRS. The IRS phase shifts in this step are kept unchanged compared to the transmission step. Finally, in the CAL step, the BS performs the localization estimation of the target based on the received echo signal and optimizes the BS transmit waveforms as well as the IRS phase shifts in the transmission and reception steps required for the $ \left( {c + 1} \right) $-th cycle.
	
	\begin{figure}[H]
		\centerline{\includegraphics[width=0.38\textwidth]{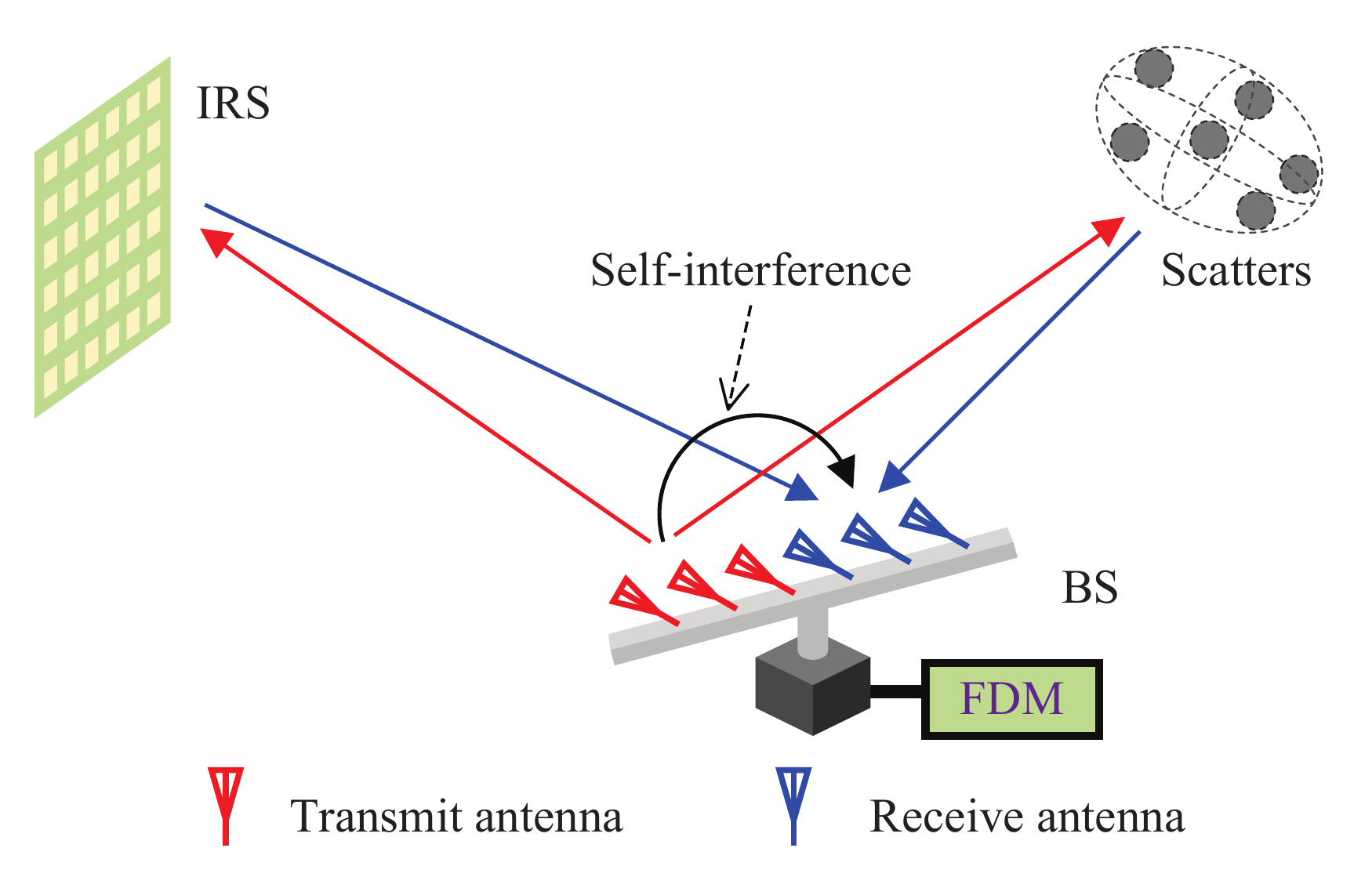}}
		\caption{Separate BS-IRS channel estimation, where the BS operates in FDM.}
		\label{fig_CE}
	\end{figure}
	
	\section{Separate BS-IRS Channel Estimation}\label{Sec_ChannelEstimation}
	In this section, we propose a separate BS-IRS channel estimation scheme, where the estimated BS-IRS CSI can be further exploited in IRS-aided wireless target localization.
	
	As shown in Fig. \ref{fig_CE}, similar to \cite{9400843_Hu}, in our proposed separate BS-IRS channel estimation scheme, part of the BS antennas first transmit pilot symbols to the IRS via the downlink channel, then the IRS reflects the pilot symbols to the remaining BS antennas through the uplink channel.
	Denote the index sets of the $ M_t $ BS transmit antennas and the rest $ \left( {{M} - {M_t}} \right) $ BS receive antennas as $ \mathcal{A} $ and $ \mathcal{B} $, respectively, then we have $ {\cal A} \cup {\cal B} = {\cal M} $, with $ {\cal M} \buildrel \Delta \over = \left\{ {1,2, \cdots ,M} \right\} $, and $ {\cal A} \cap {\cal B} = \emptyset  $. Based on this, the baseband equivalent channels from the BS transmit antennas to the IRS, from the IRS to the BS receive antennas are denoted by $ {{\bf{G}}_{\cal A}} \in {\mathbb{C}^{N \times {M_t}}} $ and $ {\bf{G}}_{\cal B}^T \in {\mathbb{C}^{\left( {{M} - {M_t}} \right) \times N}} $, respectively, where the matrices $ {{\bf{G}}_{\cal A}} $ and $ {{\bf{G}}_{\cal B}} $ represent the channel matrix $ {\bf{G}} $ divided based on the column indices of the elements into sets $ \cal A $ and $ \cal B $, respectively.
	The signal propagation paths from the transmit antennas of the BS in set $ \mathcal{A} $ to the receive antennas of the BS in set $ \mathcal{B} $ include the reflection path via the IRS $ {\bf{G}}_{\cal B}^T{\bf{\Theta }}{{\bf{G}}_{\cal A}} $, the scattering channel through the surrounding scatterers $ {{\bf{H}}_{{\cal A},{\cal B}}^{{\rm{Ref}}}} $, and the self-interference direct transmission channel $ {{\bf{H}}_{{\cal A},{\cal B}}^{{\rm{SI}}}} $.
	The proposed pilot transmission frame consists of $ P $ sub-frames, and each sub-frame lasts for $ L $ time slots. The sets of BS transmit and receive antennas remain unchanged in each sub-frame, and vary between different sub-frames. Thus, the sets $ \mathcal{A} $ and $ \mathcal{B} $ are functions of sub-frame index $ p $.
	Let $ {{\bf{x}}_{\cal A}}\left( {p,l} \right) \in {\mathbb{C}^{{M_t} \times 1}} $ be the transmitted pilot signal in the $ l $-th time slot of the $ p $-th sub-frame. Then, the received signal at the BS receive antennas is given by
	\begin{align}
	{{\bf{y}}_{\cal B}}\left( {p,l} \right) = & \ {\bf{G}}_{\cal B}^T{\bf{\Theta }}\left( {p,l} \right){{\bf{G}}_{\cal A}}{{\bf{x}}_{\cal A}}\left( {p,l} \right) \notag\\
	&+ \left( {{\bf{H}}_{{\cal A},{\cal B}}^{{\rm{Ref}}} + {\bf{H}}_{{\cal A},{\cal B}}^{{\rm{SI}}}} \right){{\bf{x}}_{\cal A}}\left( {p,l} \right) + {{\bf{n}}_{\cal B}}\left( {p,l} \right),
	\end{align}
	where $ {\bf{\Theta }}\left( {p,l} \right) $ and $ {{\bf{n}}_{\cal B}}\left( {p,l} \right) \sim {\cal C}{\cal N}\left( {{\bf{0}},\sigma^2{{\bf{I}}_{{M} - {M_t}}}} \right) $ represent the reflection coefficient matrix of the IRS and the received noise in the $ l $-th time slot of the $ p $-th sub-frame, respectively. Define $ {{{\bf{\tilde y}}}_{\cal B}}\big( {p,\tilde l} \big) \buildrel \Delta \over = {{\bf{y}}_{\cal B}}\left( {p,l + 1} \right) - {{\bf{y}}_{\cal B}}\left( {p,l} \right) $, and let $ {{\bf{x}}_{\cal A}}\left( {p,l + 1} \right) = {{\bf{x}}_{\cal A}}\left( {p,l} \right) \buildrel \Delta \over = {{\bf{x}}_{\cal A}}\big( {p,\tilde l} \big) $, we have 
	\begin{align}\label{tilde_yB}
	{{\bf{\tilde y}}_{\cal B}}\big( {p,\tilde l} \big) = {\bf{G}}_{\cal B}^T\Delta {\bf{\Theta }}\big( {p,\tilde l} \big){{\bf{G}}_{\cal A}}{{\bf{x}}_{\cal A}}\big( {p,\tilde l} \big) + \Delta {{\bf{n}}_{\cal B}}\big( {p,\tilde l} \big),
	\end{align}
	where $ \Delta {\bf{\Theta }}\big( {p,\tilde l} \big) = {\bf{\Theta }}\left( {p,l + 1} \right) - {\bf{\Theta }}\left( {p,l} \right) $ and $ \Delta {{\bf{n}}_{\cal B}}\big( {p,\tilde l} \big) = {{\bf{n}}_{\cal B}}\left( {p,l + 1} \right) - {{\bf{n}}_{\cal B}}\left( {p,l} \right) $, with $ \Delta {{\bf{n}}_{\cal B}}\big( {p,\tilde l} \big) \sim {\cal C}{\cal N}\left( {{\bf{0}},2\sigma ^2{{{\bf{I}}_{{M} - {M_t}}}}} \right) $.
	We first rewrite the $ {\bf{G}}_{\cal B}^T\Delta {\bf{\Theta }}\big( {p,\tilde l} \big){{\bf{G}}_{\cal A}}{{\bf{x}}_{\cal A}}\big( {p,\tilde l} \big) $ term in (\ref{tilde_yB}) as
	\begin{align}
	&{\bf{G}}_{\cal B}^T\Delta {\bf{\Theta }}\big( {p,\tilde l} \big){{\bf{G}}_{\cal A}}{{\bf{x}}_{\cal A}}\big( {p,\tilde l} \big) \notag\\
	&= {{\bf{I}}_{{M} - {M_t}}}{\bf{G}}_{\cal B}^T\Delta {\bf{\Theta }}\big( {p,\tilde l} \big){{\bf{G}}_{\cal A}}{{\bf{x}}_{\cal A}}\big( {p,\tilde l} \big) \notag\\
	&\mathop  = \limits^{\left( a \right)} \big( {{\bf{x}}_{\cal A}^T\big( {p,\tilde l} \big) \otimes {{\bf{I}}_{{M} - {M_t}}}} \big){\rm{vec}}\big( {{\bf{G}}_{\cal B}^T\Delta {\bf{\Theta }}\big( {p,\tilde l} \big){{\bf{G}}_{\cal A}}} \big) \notag\\
	&\mathop  = \limits^{\left( b \right)} \big( {{\bf{x}}_{\cal A}^T\big( {p,\tilde l} \big) \otimes {{\bf{I}}_{{M} - {M_t}}}} \big)\left( {{\bf{G}}_{\cal A}^T\diamond {\bf{G}}_{\cal B}^T} \right)\Delta {\bm{\theta }}\big( {p,\tilde l} \big),
	\end{align}
	where $ \Delta {\bf{\Theta }}\big( {p,\tilde l} \big) = {\rm{diag}}\big( {\Delta {\bm{\theta }}\big( {p,\tilde l} \big)} \big) $, $ \left( a \right) $ holds due to $ {\rm{vec}}\left( {{\bf{ABC}}} \right) = \left( {{{\bf{C}}^T} \otimes {\bf{A}}} \right){\rm{vec}}\left( {\bf{B}} \right) $ \cite{matrix_Zhang}, and $ \left( b \right) $ is due to $ {\rm{vec}}\left( {{\bf{A\Lambda C}}} \right) = \left( {{{\bf{C}}^T}\diamond {\bf{A}}} \right){\bm{\lambda }} $, with $ {\bf{\Lambda }} = {\rm{diag}}\left( {\bm{\lambda }} \right) $, and $ \diamond $ is the Kathri-Rao product \cite{matrix_Zhang}. Therefore, we can finally re-express $ {{\bf{\tilde y}}_{\cal B}}\big( {p,\tilde l} \big) $ as
	\begin{align}
	&{{{\bf{\tilde y}}}_{\cal B}}\big( {p,\tilde l} \big) = {{\bf{M}}_{\cal A}}\big( {p,\tilde l} \big)\left( {{\bf{G}}_{\cal A}^T\diamond {\bf{G}}_{\cal B}^T} \right)\Delta {\bm{\theta }}\big( {p,\tilde l} \big) + \Delta {{\bf{n}}_{\cal B}}\big( {p,\tilde l} \big) \notag\\
	&= \big( {\Delta {{\bm{\theta }}^T}\big( {p,\tilde l} \big) \otimes {{\bf{M}}_{\cal A}}\big( {p,\tilde l} \big)} \big){\rm{vec}}\left( {{\bf{G}}_{\cal A}^T\diamond {\bf{G}}_{\cal B}^T} \right) + \Delta {{\bf{n}}_{\cal B}}\big( {p,\tilde l} \big),
	\end{align}
	where $ {{\bf{M}}_{\cal A}}\big( {p,\tilde l} \big) \buildrel \Delta \over = {\bf{x}}_{\cal A}^T\big( {p,\tilde l} \big) \otimes {{\bf{I}}_{{M} - {M_t}}} $. By stacking the generated $ C $ vectors $ {{{\bf{\tilde y}}}_{\cal B}}\left( {p,1} \right), \cdots ,{{{\bf{\tilde y}}}_{\cal B}}\left( {p,C} \right) $ into $ {{\bf{\tilde y}}_{\cal B}}\left( p \right) = {\left[ {{\bf{\tilde y}}_{\cal B}^T\left( {p,1} \right), \cdots ,{\bf{\tilde y}}_{\cal B}^T\left( {p,C} \right)} \right]^T} $, we have
	\begin{align}\label{vec_y}
	{{{\bf{\tilde y}}}_{\cal B}}\left( p \right) = {{\bf{\Phi }}^{\left( p \right)}}{{\bm{\omega }}^{\left( p \right)}} + \Delta {{\bf{n}}_{\cal B}}\left( p \right),
	\end{align}
	where $ {{\bf{\Phi }}^{\left( p \right)}} \buildrel \Delta \over = \left[ {\begin{array}{*{20}{c}}
		{\Delta {{\bm{\theta }}^T}\left( {p,1} \right) \otimes {{\bf{M}}_{\cal A}}\left( {p,1} \right)}\\
		\vdots \\
		{\Delta {{\bm{\theta }}^T}\left( {p,C} \right) \otimes {{\bf{M}}_{\cal A}}\left( {p,C} \right)}
		\end{array}} \right] $, $ {{\bm{\omega }}^{\left( p \right)}} \buildrel \Delta \over = {\rm{vec}}\big( {{\bf{G}}_{{\cal A}\left( p \right)}^T\diamond {\bf{G}}_{{\cal B}\left( p \right)}^T} \big) $, and $ \Delta {{\bf{n}}_{\cal B}}\left( p \right) = {\left[ {\Delta {\bf{n}}_{\cal B}^T\left( {p,1} \right), \cdots ,\Delta {\bf{n}}_{\cal B}^T\left( {p,C} \right)} \right]^T} $. By applying the least squares (LS) estimator, the estimated $ {{\bm{\omega }}^{\left( p \right)}} $ is given by \cite{1597555_Biguesh}
	\begin{align}\label{omega_esti}
	{{{\bm{\hat \omega }}}^{\left( p \right)}} = {{\bf{\Phi }}^{\left( p \right),\dag }}{{{\bf{\tilde y}}}_{\cal B}}\left( p \right),
	\end{align}
	where $ {{\bf{\Phi }}^{\left( p \right),\dag }} = {\left( {{{\bf{\Phi }}^{\left( p \right),H}}{{\bf{\Phi }}^{\left( p \right)}}} \right)^{ - 1}}{{\bf{\Phi }}^{\left( p \right),H}} $ is the pseudoinverse of $ {{{\bf{\Phi }}^{\left( p \right)}}} $. Since $ {{\bf{\Phi }}^{\left( p \right)}} \in {\mathbb{C}^{C\left( {{M} - {M_t}} \right) \times N{M_t}\left( {{M} - {M_t}} \right)}} $, it can be readily seen that $ C $ should satisfy $ C \ge N{M_t} $. From $ {{{\bm{\hat \omega }}}^{\left( p \right)}} $, we can obtain an estimate of $ {\left\{ {{g_{n,a}}{g_{n,b}}} \right\}_{n \in {\cal N},a \in {\cal A},b \in {\cal B}}} $, where $ {g_{n,a}}, \forall n \in \mathcal{N}, a \in \mathcal{M} $, is the $ \left( {n,a} \right) $-th element of $ {\bf{G}} $. By choosing different sets $ \mathcal{A} $ and $ \mathcal{B} $, we can get several estimates of $ {g_{n,a}}{g_{n,b}},\forall n \in {\cal N},a \in {\cal M},b \in {\cal M}\backslash \left\{ a \right\} $, and further derive an estimate for $ {g_{n,a}},\forall n \in {\cal N},a \in {\cal M} $, which will be given in detail later. In general, the accuracy of the estimated value of $ {g_{n,a}} $ increases with the number of the estimated value of $ {g_{n,a}}{g_{n,b}},\forall n \in {\cal N},a \in {\cal M},b \in {\cal M}\backslash \left\{ a \right\} $. Based on this, we express the efficiency of channel estimation in terms of the obtained number of the estimated $ {g_{n,a}}{g_{n,b}},\forall n \in {\cal N},a \in {\cal M},b \in {\cal M}\backslash \left\{ a \right\} $, per pilot overhead. The following proposition gives the number of transmit antennas when there is no prior knowledge of the estimated channel $ {\bf{G}} $, thus maximizing the efficiency of channel estimation.
	
	\newtheorem{theorem}{Proposition}
	\begin{theorem}
		When there is no prior knowledge of the estimated channel $ {\bf{G}} $, the efficiency of channel estimation is highest when the number of transmit antennas is set to 1.
	\end{theorem}
	
	\begin{IEEEproof}
		According to the definition of $ {\bm{\omega }} $, it can be readily seen that the estimate of one element in $ {\bf{G}} $ is only related to the other elements in the row where it is located. Therefore, we take the estimate of the elements in the first row of $ {\bf{G}} $ as an example. At the cost of the pilot overhead $ N{M_t} $, the number of estimate of $ {g_{1,a}}{g_{1,b}},\forall a \in {\cal A},b \in {\cal B} $, used for estimating the elements in the first row of $ {\bf{G}} $ is $ {M_t}\left( {{M} - {M_t}} \right) $. For the sake of fairness, for any $ a \in \mathcal{M} $ and $ b \in {\cal M}\backslash \left\{ a \right\} $, the number of estimates for different $ {g_{1,a}}{g_{1,b}} $ should be the same. Therefore, we need to conduct a total number of $ C\left( {M,{M_t}} \right)  = \frac{{M!}}{{{M_t}!\left( {M - {M_t}} \right)!}} $ 
		different transmit/receive antenna combinations. Accordingly, when the number of the transmit antennas is set to $ M_t $, the efficiency of channel estimation is calculated as
		\begin{align}
		\eta  &= \frac{{{M_t}\left( {{M} - {M_t}} \right)C\left( {M,{M_t}} \right)}}{{{M}\left( {{M} - 1} \right)/2}}\frac{1}{{C\left( {M,{M_t}} \right)N{M_t}}} \notag\\
		&= \frac{{2\left( {{M} - {M_t}} \right)}}{{N{M}\left( {{M} - 1} \right)}}.
		\end{align}
		It can be seen that $ \eta $ decreases monotonically with $ M_t $. Thus, we conclude that the efficiency of channel estimation is highest when the number of transmit antennas is set to 1.
	\end{IEEEproof}
	
	According to (\ref{vec_y}) and (\ref{omega_esti}), we have
	\begin{align}
	{{{\bm{\hat \omega }}}^{\left( p \right)}} = {{\bm{\omega }}^{\left( p \right)}} + {\big( {{{\bf{\Phi }}^{\left( p \right),H}}{{\bf{\Phi }}^{\left( p \right)}}} \big)^{ - 1}}{{\bf{\Phi }}^{\left( p \right),H}}\Delta {{\bf{n}}_{\cal B}}\left( p \right),
	\end{align}
	which shows that $ {{{\bm{\hat \omega }}}^{\left( p \right)}} $ is a circularly symmetric Gaussian random vector whose mean and covariance matrix are respectively given by
	\begin{align}\label{Expectation}
	\mathbb{E}\big\{ {{{{\bm{\hat \omega }}}^{\left( p \right)}}} \big\} = {{\bm{\omega }}^{\left( p \right)}},
	\end{align}
	and
	\begin{align}\label{Covariance}
	{{\bf{R}}_{{{{\bm{\hat \omega }}}^{\left( p \right)}}}} &= \big\{ {\big( {{{{\bm{\hat \omega }}}^{\left( p \right)}} - {{\bm{\omega }}^{\left( p \right)}}} \big){{\big( {{{{\bm{\hat \omega }}}^{\left( p \right)}} - {{\bm{\omega }}^{\left( p \right)}}} \big)}^H}} \big\} \notag\\
	&= 2\sigma^2{\big( {{{\bf{\Phi }}^{\left( p \right),H}}{{\bf{\Phi }}^{\left( p \right)}}} \big)^{ - 1}}.
	\end{align}
	Based on (\ref{Expectation}) and (\ref{Covariance}), we have
	\begin{align}
	{{{\bm{\hat \omega }}}^{\left( p \right)}} \sim {\cal C}{\cal N}\big( {{{\bm{\omega }}^{\left( p \right)}},{{\bf{R}}_{{{{\bm{\hat \omega }}}^{\left( p \right)}}}}} \big).
	\end{align}
	Given $ {{{\bm{\omega }}^{\left( p \right)}}} $,  the probability density function (pdf) of the observed vector $ {{{{\bm{\hat \omega }}}^{\left( p \right)}}} $ is expressed as
	\begin{align}
	&f\big( {{{{\bm{\hat \omega }}}^{\left( p \right)}}|{{\bm{\omega }}^{\left( p \right)}}} \big) = \frac{1}{{{\pi ^{N{M_t}\left( {{M} - {M_t}} \right)}}\det {{\bf{R}}_{{{{\bm{\hat \omega }}}^{\left( p \right)}}}}}} \times \notag\\
	& \qquad \exp \left( { - {{\big( {{{{\bm{\hat \omega }}}^{\left( p \right)}} - {{\bm{\omega }}^{\left( p \right)}}} \big)}^H}{\bf{R}}_{{{{\bm{\hat \omega }}}^{\left( p \right)}}}^{ - 1}\big( {{{{\bm{\hat \omega }}}^{\left( p \right)}} - {{\bm{\omega }}^{\left( p \right)}}} \big)} \right).
	\end{align}
	For all the generated observation vectors $ \left\{ {{{{\bm{\hat \omega }}}^{\left( p \right)}}} \right\}_{p = 1}^P $, where $ P = C\left( {M,{M_t}} \right) $,
	the corresponding maximum likelihood (ML) estimator is given by
	\begin{align}\label{ML_esti}
	\left\{ {{{\hat g}_{n,a}}} \right\} = \arg \mathop {\max }\limits_{\left\{ {{g_{n,a}}} \right\}} {\rm{ }}\sum\limits_p {\log f\big( {{{{\bm{\hat \omega }}}^{\left( p \right)}}|{{\bm{\omega }}^{\left( p \right)}}} \big)}.
	\end{align}
	Since $ {{\bf{R}}_{{{{\bm{\hat \omega }}}^{\left( p \right)}}}} $ is only related to $ {\bf{\Theta }}\left( {p,l} \right) $ and $ {{\bf{x}}_{\cal A}}\left( {p,l} \right), \forall l $, from (\ref{Covariance}), we set $ {\bf{\Theta }}\left( {1,l} \right) =  \cdots  = {\bf{\Theta }}\left( {P,l} \right) $ and $ {{\bf{x}}_{\cal A}}\left( {1,l} \right) =  \cdots  = {{\bf{x}}_{\cal A}}\left( {P,l} \right) $ such that $ {{\bf{R}}_{{{{\bm{\hat \omega }}}^{\left( p \right)}}}} = {{\bf{R}}_{{\bm{\hat \omega }}}},\forall p $, for simplicity. Accordingly, by neglecting the irrelevant constant terms, problem (\ref{ML_esti}) is changed into
	\begin{align}\label{ML_esti_2}
	\left\{ {{{\hat g}_{n,a}}} \right\} = &\arg \mathop {\min }\limits_{\left\{ {{g_{n,a}}} \right\}} {\rm{ }} J\big( {\big\{ {{{\bm{\omega }}^{\left( p \right)}}} \big\}} \big) \buildrel \Delta \over = \notag\\
	&\quad \sum\limits_p {{{\big( {{{{\bm{\hat \omega }}}^{\left( p \right)}} - {{\bm{\omega }}^{\left( p \right)}}} \big)}^H}{\bf{R}}_{{\bm{\hat \omega }}}^{ - 1}\big( {{{{\bm{\hat \omega }}}^{\left( p \right)}} - {{\bm{\omega }}^{\left( p \right)}}} \big)}.
	\end{align}
	
	\subsection{Element-Wise Optimization \& Refinement}
	It can be readily seen that for each $ {g_{n,a}} $, the function $ J\left( {\left\{ {{{\bm{\omega }}^{\left( p \right)}}} \right\}} \right) $ is a quadratic function with respect to it with the other elements in $ {\bf{G}} $ fixed. Based on this, we propose an iterative channel estimation algorithm that corrects the estimate of the channel matrix $ {\bf{G}} $ in an element-wise manner. Specifically, the estimates of all the elements in $ {\bf{G}} $ are refined one-by-one by minimizing the function $ J\left( {\left\{ {{{{\bm{\hat \omega }}}^{\left( p \right)}}} \right\}} \right) $ with the other elements fixed. The above procedure is repeated until the estimate of $ {\bf{G}} $ converges. The details are given as follows.
	
	To be specific, in the $ i $-th iteration, we fix the estimates of other $ \left( N{M_t} - 1 \right) $ channel coefficients, namely, $ \hat g_{1,1}^{\left( i \right)}, \cdots ,\hat g_{1,{M_t}}^{\left( i \right)}, \cdots ,\hat g_{n,a - 1}^{\left( i \right)},\hat g_{n,a + 1}^{\left( {i - 1} \right)}, \cdots ,\hat g_{N,1}^{\left( {i - 1} \right)}, \cdots \hat g_{N,{M_t}}^{\left( {i - 1} \right)} $, and refine the estimate of $ {{\hat g}_{n,a}} $, which means that $ {{\hat g}_{1,1}}, \cdots ,{{\hat g}_{1,{M_t}}}, \cdots ,{{\hat g}_{n,a - 1}} $ have been refined before $ {{\hat g}_{n,a}} $ in the current $ i $-th iteration, and $ {{\hat g}_{n,a + 1}}, \cdots ,{{\hat g}_{N,1}}, \cdots {{\hat g}_{N,{M_t}}} $ have not been refined in the current iteration. Thereby, the sub-problem of refining the estimate of $ {g_{n,a}} $ can be reformulated as
	\begin{align}\label{opti_g_{n,a}}
	\hat g_{n,a}^{\left( i \right)} =& \arg \mathop {\min }\limits_{{g_{n,a}}} {\rm{ }}J\left( {{g_{n,a}}|g_{1,1}^{\left( i \right)}, \cdots ,g_{1,{M_t}}^{\left( i \right)},} \right. \notag\\
	&\quad \left. { \cdots ,g_{n,a - 1}^{\left( i \right)},g_{n,a + 1}^{\left( {i - 1} \right)}, \cdots ,g_{N,1}^{\left( {i - 1} \right)}, \cdots g_{N,{M_t}}^{\left( {i - 1} \right)}} \right).
	\end{align}
	The closed-form optimal solution to problem (\ref{opti_g_{n,a}}) can be obtained by taking the first-order derivative of the function $ J $ with respect to $ g_{n,a}^ *  $ and setting it to zero.
	The derivative of the function $ J $ with respect to $ g_{n,a}^ *  $ can be expressed as
	\begin{align}\label{Partial_J}
	\frac{{\partial J}}{{\partial g_{n,a}^ * }} = \sum\limits_p {{{\left( {\frac{{\partial J}}{{\partial {{\bm{\omega }}^{\left( p \right), * }}}}} \right)}^T}\frac{{\partial {{\bm{\omega }}^{\left( p \right), * }}}}{{\partial g_{n,a}^ * }}},
	\end{align}
	where $ \frac{{\partial J}}{{\partial {{\bm{\omega }}^{\left( p \right), * }}}} $ is given by
	\begin{align}\label{Partial_J_omega}
	\frac{{\partial J}}{{\partial {{\bm{\omega }}^{\left( p \right), * }}}} = {\bf{R}}_{{\bm{\hat \omega }}}^{ - 1}\big( {{{{\bm{\hat \omega }}}^{\left( p \right)}} - {\rm{vec}}\big( {{\bf{G}}_{{\cal A}\left( p \right)}^T\diamond {\bf{G}}_{{\cal B}\left( p \right)}^T} \big)} \big).
	\end{align}
	We first define a function $ {\chi _\alpha }\left(  \cdot  \right) $ as
	\begin{align}
	{\bf{Y}} = {\chi _\alpha }\left( {\bf{X}} \right),{\bf{Y}}\left( {m,n} \right) = \left\{ {\begin{array}{*{20}{c}}
		{1,{\text{ if }}{\bf{X}}\left( {m,n} \right) \text{ is } \alpha ,}\\
		{0,{\text{ otherwise,}}}
		\end{array}} \right.
	\end{align}
	where $ \alpha $ is a variable. Therefore, we can reformulate the term $ {\bf{G}}_{{\cal A}\left( p \right)}^T\diamond {\bf{G}}_{{\cal B}\left( p \right)}^T $ as
	\begin{align}\label{G_khatri-rao}
	&{\bf{G}}_{{\cal A}\left( p \right)}^T\diamond {\bf{G}}_{{\cal B}\left( p \right)}^T \notag\\
	&= \big( {{\bf{G}}_{{\cal A}\left( p \right)}^T - {g_{n,a}}{\chi _{{g_{n,a}}}}\big( {{\bf{G}}_{{\cal A}\left( p \right)}^T} \big) + {g_{n,a}}{\chi _{{g_{n,a}}}}\big( {{\bf{G}}_{{\cal A}\left( p \right)}^T} \big)} \big) \notag\\
	&\quad \diamond \big( {{\bf{G}}_{{\cal B}\left( p \right)}^T - {g_{n,a}}{\chi _{{g_{n,a}}}}\big( {{\bf{G}}_{{\cal B}\left( p \right)}^T} \big) + {g_{n,a}}{\chi _{{g_{n,a}}}}\big( {{\bf{G}}_{{\cal B}\left( p \right)}^T} \big)} \big) \notag\\
	& = {\bf{\Xi }}_{{g_{n,a}},1}^{\left( p \right)} + {g_{n,a}}{\bf{\Xi }}_{{g_{n,a}},2}^{\left( p \right)},
	\end{align}
	where
	\begin{align}
	{\bf{\Xi }}_{{g_{n,a}},1}^{\left( p \right)}
	= \ & \left( {{\bf{G}}_{{\cal A}\left( p \right)}^T - {g_{n,a}}{\chi _{{g_{n,a}}}}\big( {{\bf{G}}_{{\cal A}\left( p \right)}^T} \big)} \right) \notag\\
	&\diamond \left( {{\bf{G}}_{{\cal B}\left( p \right)}^T - {g_{n,a}}{\chi _{{g_{n,a}}}}\big( {{\bf{G}}_{{\cal B}\left( p \right)}^T} \big)} \right),
	\end{align}
	and
	\begin{align}
	&{\bf{\Xi }}_{{g_{n,a}},2}^{\left( p \right)} \notag \\
	& =  {\chi _{{g_{n,a}}}}\big( {{\bf{G}}_{{\cal A}\left( p \right)}^T} \big)\diamond \big( {{\bf{G}}_{{\cal B}\left( p \right)}^T - {g_{n,a}}{\chi _{{g_{n,a}}}}\big( {{\bf{G}}_{{\cal B}\left( p \right)}^T} \big)} \big) \notag\\
	& \quad + \big( {{\bf{G}}_{{\cal A}\left( p \right)}^T - {g_{n,a}}{\chi _{{g_{n,a}}}}\big( {{\bf{G}}_{{\cal A}\left( p \right)}^T} \big)} \big)\diamond {\chi _{{g_{n,a}}}}\big( {{\bf{G}}_{{\cal B}\left( p \right)}^T} \big).
	\end{align}
	We can see that neither $ {\bf{\Xi }}_{{g_{n,a}},1}^{\left( p \right)} $ nor $ {\bf{\Xi }}_{{g_{n,a}},2}^{\left( p \right)} $ contains $ {{g_{n,a}}} $.
	Substituting (\ref{Partial_J_omega}) and (\ref{G_khatri-rao}) into (\ref{Partial_J}), we have
	\begin{align}
	\frac{{\partial J}}{{\partial g_{n,a}^ * }} = \ & \sum\limits_p {{{\big( {{{{\bm{\hat \omega }}}^{\left( p \right)}} - {\rm{vec}}\big( {{\bf{\Xi }}_{{g_{n,a}},1}^{\left( p \right)}} \big)} \big)}^T}{\bf{R}}_{{\bm{\hat \omega }}}^{ - T}\frac{{\partial {{\bm{\omega }}^{\left( p \right), * }}}}{{\partial g_{n,a}^ * }}} \notag\\
	&- {g_{n,a}}\sum\limits_p {{\rm{ve}}{{\rm{c}}^T}\big( {{\bf{\Xi }}_{{g_{n,a}},2}^{\left( p \right)}} \big){\bf{R}}_{{\bm{\hat \omega }}}^{ - T}\frac{{\partial {{\bm{\omega }}^{\left( p \right), * }}}}{{\partial g_{n,a}^ * }}},
	\end{align}
	where the term $ \frac{{\partial {{\bm{\omega }}^{\left( p \right), * }}}}{{\partial g_{n,a}^ * }} $ can be expressed as
	\begin{align}\label{Partial_omega}
	\frac{{\partial {{\bm{\omega }}^{\left( p \right), * }}}}{{\partial g_{n,a}^ * }} = \ & {\rm{ve}}{{\rm{c}}^ * }\left( {{\bf{G}}_{\cal A}^T\diamond \left( {{\chi _{{g_{n,a}}}}\left( {{\bf{G}}_{\cal B}^T} \right)} \right)} \right) \notag \\
	&+ {\rm{ve}}{{\rm{c}}^ * }\left( {\left( {{\chi _{{g_{n,a}}}}\left( {{\bf{G}}_{\cal A}^T} \right)} \right)\diamond {\bf{G}}_{\cal B}^T} \right).
	\end{align}
	Therefore, by setting $ \frac{{\partial J}}{{\partial g_{n,a}^ * }} = 0 $, the optimal $ {g_{n,a}} $ is given by
	\begin{align}
	g_{n,a}^{{\rm{opti}}} = \frac{{\sum\limits_p {{{\big( {{{{\bm{\hat \omega }}}^{\left( p \right)}} - {\rm{vec}}\big( {{\bf{\Xi }}_{{g_{n,a}},1}^{\left( p \right)}} \big)} \big)}^T}{\bf{R}}_{{\bm{\hat \omega }}}^{ - T}\frac{{\partial {{\bm{\omega }}^{\left( p \right), * }}}}{{\partial g_{n,a}^ * }}} }}{{\sum\limits_p {{\rm{ve}}{{\rm{c}}^T}\big( {{\bf{\Xi }}_{{g_{n,a}},2}^{\left( p \right)}} \big){\bf{R}}_{{\bm{\hat \omega }}}^{ - T}\frac{{\partial {{\bm{\omega }}^{\left( p \right), * }}}}{{\partial g_{n,a}^ * }}} }}.
	\end{align}
	
	\subsection{Initialization of Iterative Channel Estimation Algorithm}
	A good initialization of the iterative channel estimation algorithm helps to improve the final performance and reduce the computational cost.
	In this subsection, we present how to calculate the initial channel estimate for problem (\ref{ML_esti_2}).
	Specifically, we first average all the obtained estimates of $ {g_{n,a}}{g_{n,b}},\forall n \in {\cal N},a \in {\cal M},b \in {\cal M}\backslash \left\{ a \right\} $, and denote the mean value as $ {{\bar h}_{n,a,b}} $.
	For each tuple $ \left( {p,q} \right) $, where $ p \ne l $, $ q \ne l $, and $ p \ne q $, we can obtain an estimate of $ {g_{n,l}} $ as
	\begin{align}\label{esti_g_n,1}
	\hat g_{n,l}^{\left( {p,q} \right)} = \sqrt {\frac{{{{\bar h}_{n,l,p}}{{\bar h}_{n,l,q}}}}{{{{\bar h}_{n,p,q}}}}}. 
	\end{align}
	The initial estimate of $ \hat g_{n,l}^{\left( 0 \right)} $ is then given by the geometric mean of all the obtained $ \big\{ {\hat g_{n,l}^{\left( {p,q} \right)}} \big\} $, i.e.,
	\begin{align}\label{esti_g_n,1_(0)}
	\hat g_{n,l}^{\left( 0 \right)} = {\Bigg( {\prod\limits_{\left( {p,q} \right)} {\hat g_{n,l}^{\left( {p,q} \right)}} } \Bigg)^{1/{\rm{card}}\left( {\left\{ {\left( {p,q} \right)} \right\}} \right)}}.
	\end{align}
	With the determined $ \hat g_{n,l}^{\left( 0 \right)} $, the initial estimates of $ \hat g_{n,1}^{\left( 0 \right)}, \cdots ,\hat g_{n,l - 1}^{\left( 0 \right)},\hat g_{n,l + 1}^{\left( 0 \right)}, \cdots ,\hat g_{n,{M_t}}^{\left( 0 \right)} $ are calculated by
	\begin{align}\label{esti_g_n,a_(0)}
	\hat g_{n,a}^{\left( 0 \right)} = \frac{{{{\bar h}_{n,l,a}}}}{{\hat g_{n,l}^{\left( 0 \right)}}},a = 1, \cdots ,l - 1,l + 1, \cdots {M_t},n \in {\cal N}.
	\end{align}
	
	\subsection{Sign Ambiguity of Estimated Channel Coefficients}
	We can see that there are two values of $ {\hat g_{n,l}^{\left( {p,q} \right)}} $ that satisfy (\ref{esti_g_n,1}), corresponding to the positive and negative real parts, respectively. This leads to $ {\bf{\hat g}}_n^{T,\left( 0 \right)} \buildrel \Delta \over = \big[ {\hat g_{n,1}^{\left( 0 \right)},\hat g_{n,2}^{\left( 0 \right)}, \cdots ,\hat g_{n,{M_t}}^{\left( 0 \right)}} \big], \forall n \in \mathcal{N} $, being independently and randomly picked from two vectors with opposite signs according to (\ref{esti_g_n,1_(0)}) and (\ref{esti_g_n,a_(0)}). Therefore, the final estimated channel should be
	\begin{align}\label{G_esti_final}
	{{{\bf{\hat G}}}^{{\text{final}}}} = {\rm{diag}}\left( {\bm{\delta }} \right){\bf{\hat G}},
	\end{align}
	where $ {{\bf{\hat G}}} $ is obtained by solving problem (\ref{ML_esti_2}), and $ {\bm{\delta }}\left( n \right), \forall n \in \mathcal{N} $, is a randomly selected value from $ \left\{ { - 1,1} \right\} $ to be determined.
	
	\emph{Remark 1:} Adopting our proposed separate BS-IRS channel estimation method, we reduce the $ NM_t $ continuously valued quantities to be estimated, i.e., $ \bf{G} $, to $ N $ binary discretely valued quantities to be estimated, i.e., $ \bm{\delta} $, thus greatly alleviating the subsequent channel estimation difficulty.
	
	\section{Proposed Target Localization Scheme}
	In this section, we present in detail the target localization scheme using the partial information of the BS-IRS channel estimated in Section \ref{Sec_ChannelEstimation}. The complete CSI of the BS-IRS link can be simultaneously obtained with the target localization.
	
	\subsection{Signal Transmission Model}
	Let $ {{\bf{X}}^c} = \left[ {{{\bf{x}}^c}, \cdots ,{{\bf{x}}^c}} \right] \in {\mathbb{C}^{M \times L}} $ denote the waveform matrix generated by the BS in the $ c $-th target localization cycle, with $ L $ being the number of snapshots.
	In the transmission step, the waveforms $ {{\bf{X}}^c} $ emitted by the BS pass through the IRS to reach the target. While in the reception step, it is then scattered by the target and finally listened to by the BS via the IRS. 
	It is worth noting that to reduce the complexity of the hardware operation, the IRS reflection coefficient matrix is set to be the same in both transmission and reception steps, which is different from the work in~\cite{9725255_Zhang} where the IRS phase shifts are set separately in the two steps.
	Therefore, in combination with (\ref{TargetRes_matrix}), the echo signals received by the BS in the $ c $-th target localization cycle can be expressed as
	\begin{align}
	{{\bf{Y}}^c} = \alpha {{\bf{G}}^T}{{\bf{\Theta }}^{c}}{\bf{a}}{{\bf{a}}^T}{{\bf{\Theta }}^{c}}{\bf{G}}{{\bf{X}}^c} + {{\bf{N}}^c},
	\end{align}
	where $ {{\bf{N}}^c} \in {\mathbb{C}^{M \times L}} $ is the noise from the environment, and we assume that all columns of $ {{\bf{N}}^c} $ follow independent and identically distributed circularly symmetric complex Gaussian distribution with zero mean and covariance matrix $ {\sigma ^2}{{\bf{I}}_{M \times M}} $. We define $ {{\bf{y}}^c} = {\rm{vec}}\left( {{{\bf{Y}}^c}} \right) $ as the received signal vector. Since
	\begin{align}
	&{\rm{vec}}\left( {{{\bf{G}}^T}{{\bf{\Theta }}^{c}}{\bf{a}}{{\bf{a}}^T}{{\bf{\Theta }}^{c}}{\bf{G}}{{\bf{X}}^c}} \right) \notag\\
	&= {\rm{vec}}\left( {\left( {1 \otimes \left( {{{\bf{G}}^T}{{\bf{\Theta }}^{c}}{\bf{a}}{{\bf{a}}^T}{{\bf{\Theta }}^{c}}{\bf{G}}} \right)} \right)\left( {{\bf{1}}_L^T \otimes {{\bf{x}}^c}} \right)} \right) \notag\\
	&\mathop  = \limits^{\left( a \right)} {\rm{vec}}\left( {{\bf{1}}_L^T \otimes \left( {{{\bf{G}}^T}{{\bf{\Theta }}^{c}}{\bf{a}}{{\bf{a}}^T}{{\bf{\Theta }}^{c}}{\bf{G}}{{\bf{x}}^c}} \right)} \right) \notag\\
	&= {{\bf{a}}^T}{{\bf{\Theta }}^{c}}{\bf{G}}{{\bf{x}}^c}{\rm{vec}}\left( {{\bf{1}}_L^T \otimes \left( {{{\bf{G}}^T}{{\bf{\Theta }}^{c}}{\bf{a}}} \right)} \right) \notag\\
	&\mathop  = \limits^{\left( b \right)} {{\bf{a}}^T}{{\bf{\Theta }}^{c}}{\bf{G}}{{\bf{x}}^c}{\rm{vec}}\left( {{{\bf{G}}^T}{{\bf{\Theta }}^{c}}\left( {{\bf{1}}_L^T \otimes {\bf{a}}} \right)} \right),
	\end{align}
	where $ {\left( a \right)} $ and $ {\left( b \right)} $ are due to $ \left( {{\bf{AB}}} \right) \otimes \left( {{\bf{CD}}} \right) = \left( {{\bf{A}} \otimes {\bf{C}}} \right)\left( {{\bf{B}} \otimes {\bf{D}}} \right) $ \cite{matrix_Zhang}, we can rewrite $ {{\bf{y}}^c} $ as
	\begin{align}\label{y^c}
	{{\bf{y}}^c} = \gamma {{\bf{f}}^c} + {{\bf{n}}^c},
	\end{align}
	where
	$ \gamma  = \alpha {{\bf{a}}^T}{{\bf{\Theta }}^{c}}{\bf{G}}{{\bf{x}}^c} $, $ {{\bf{f}}^c} = {\rm{vec}}\left( {{{\bf{G}}^T}{{\bf{\Theta }}^{c}}\left( {{\bf{1}}_L^T \otimes {\bf{a}}} \right)} \right) $, and $ {{\bf{n}}^c} = {\rm{vec}}\left( {{{\bf{N}}^c}} \right) $.
	
	\subsection{Hypotheses Testing Framework}
	We use hypothesis testing to perform target localization. Specifically, we first form multiple hypotheses to represent different target localization results. At the end of each cycle of target localization, we update the probability of each hypothesis based on the received echo signals at the BS. When the target localization procedure terminates, the hypothesis with the highest probability will be selected as the localization result.
	Specifically, we discretize the SoI into $ I $ angular grids and denote the set of all the angular grids by $ {\cal I} \buildrel \Delta \over = \left\{ {1,2, \cdots ,I} \right\} $. Moreover, we introduce the variable $ {i_T},\forall i \in {\cal I} $, to indicate that the target is in the $ i $-th angular grid. Denote the a prior probability of hypothesis $ {i_T} $ in the $ c $-th cycle as $ {p^c}\left( {{i_T}} \right) $. According to Bayes’ theorem, the a posteriori probability of this hypothesis in the $ c $-th cycle can then be updated by the received echo signals at the BS in this cycle and used as the a priori probability in the $ \left( {c + 1} \right) $-th cycle, which can be expressed as \cite{9772371_Teng,Probability_2002}
	\begin{align}
	{p^{c + 1}}\left( {{i_T}} \right) = \frac{{{p^c}\left( {{i_T}} \right){p^c}\left( {{{\bf{y}}^c}|{i_T}} \right)}}{{\sum\nolimits_{j = 1}^I {{p^c}\left( {{j_T}} \right){p^c}\left( {{{\bf{y}}^c}|{j_T}} \right)} }},
	\end{align}
	where $ {p^c}\left( {{{\bf{y}}^c}|{j_T}} \right) $ denotes the probability to receive $ {{\bf{y}}^c} $ under the hypothesis $ {j_T} $, which is given by
	\begin{align}
	{p^c}\left( {{{\bf{y}}^c}|{j_T}} \right) = \frac{1}{{{{\left( {\pi {\sigma ^2}} \right)}^{ML}}}}\exp \left( { - \frac{{{{\big\| {{{\bf{y}}^c} - {{{\bf{\bar y}}}^c}\big( {{j_T},{{{\bm{\hat \delta }}}^c},{{\hat \gamma }^c}} \big)} \big\|}^2}}}{{{\sigma ^2}}}} \right),
	\end{align}
	where $ {{{{\bm{\hat \delta }}}^c}}\left( {{j_T}} \right) $ and $ {{{\hat \gamma }^c}}\left( {{j_T}} \right) $ denote the estimated $ {\bm{\delta }} $ and $ \gamma $ under the hypothesis $ {j_T} $ in the $ c $-th cycle, respectively, and $ {{{{\bf{\bar y}}}^c}\big( {{j_T},{{{\bm{\hat \delta }}}^c},{{\hat \gamma }^c}} \big)} $ represents the expectation of the signals received at the BS given $ {{{{\bm{\hat \delta }}}^c}} $ and $ {{{\hat \gamma }^c}} $ under the hypothesis $ j_T $ in the $ c $-th cycle, which can be calculated according to (\ref{G_esti_final}) and (\ref{y^c}) as
	\begin{align}
	{{{\bf{\bar y}}}^c}\big( {{j_T},{{{\bm{\hat \delta }}}^c},{{\hat \gamma }^c}} \big) &= {{\hat \gamma }^c}{\rm{vec}}\left( {{{{\bf{\hat G}}}^T}{\rm{diag}}\big( {{{{\bm{\hat \delta }}}^c}} \big){{\bf{\Theta }}^c}\left( {{\bf{1}}_L^T \otimes {\bf{a}}\left( {{j_T}} \right)} \right)} \right) \notag\\
	&\mathop  = \limits^{\left( a \right)} {{\hat \gamma }^c}\left( {{{\left( {{{\bf{\Theta }}^c}\left( {{\bf{1}}_L^T \otimes {\bf{a}}\left( {{j_T}} \right)} \right)} \right)}^T}\diamond {{{\bf{\hat G}}}^T}} \right){{{\bm{\hat \delta }}}^c},
	\end{align}
	where $ {\left( a \right)} $ is due to $ {\rm{vec}}\left( {{\bf{A\Lambda C}}} \right) = \left( {{{\bf{C}}^T}\diamond {\bf{A}}} \right){\bm{\lambda }} $, with $ {\bf{\Lambda }} = {\rm{diag}}\left( {\bm{\lambda }} \right) $ \cite{matrix_Zhang}. And $ {{{\bm{\hat \delta }}}^c} $ and $ {{\hat \gamma }^c} $ can be jointly estimated by applying the ML estimation method, i.e.,
	\begin{align}\label{P_jointML}
	\big\{ {{{\hat \gamma }^c},{{{\bm{\hat \delta }}}^c}} \big\} = \arg \mathop {\max }\limits_{\gamma ,{\bm{\delta }}:{\bm{\delta }}\left( n \right) \in \left\{ { - 1,1} \right\},\forall n } {\rm{ }}{p^c}\left( {{{\bf{y}}^c}|{j_T},\gamma ,{\bm{\delta }}} \right),
	\end{align}
	where
	\begin{align}
	{p^c}\left( {{{\bf{y}}^c}|{j_T},\gamma ,{\bm{\delta }}} \right) = \frac{1}{{{{\left( {\pi {\sigma ^2}} \right)}^{ML}}}}\exp \left( { - \frac{{{{\left\| {{{\bf{y}}^c} - {{{\bf{\bar y}}}^c}\left( {{j_T},{\bm{\delta }},\gamma } \right)} \right\|}^2}}}{{{\sigma ^2}}}} \right).
	\end{align}
	By neglecting the irrelevant terms, the problem in (\ref{P_jointML}) can be simplified as
	\begin{subequations}\label{P_jointML_1}
		\begin{align}
		\mathop {\min }\limits_{ \gamma ,{{{\bm{ \delta }}}}} & \ {\rm{ }}{\big\| {{{\bf{y}}^c} - \gamma {{\bf{\Phi }}^c}{{{\bm{ \delta }}}}} \big\|^2} \\
		{\rm{s}}{\rm{.t}}{\rm{. }} & \ {{\bm{ \delta }}}\left( n \right) \in \left\{ { - 1,1} \right\},\forall n \in {\cal N},
		\end{align}
	\end{subequations}
	where $ {{\bf{\Phi }}^c} = {\left( {{{\bf{\Theta }}^{c}}\left( {{\bf{1}}_L^T \otimes {\bf{a}}\left( {{j_T}} \right)} \right)} \right)^T}\diamond {{{\bf{\hat G}}}^T} $.
	Obviously, for any given $ {{\bm{ \delta }}} $, the optimal $ \gamma $ is calculated as
	\begin{align}\label{Opti_gama}
	{{ \gamma }^ \text{opt}} = \frac{{{{{\bm{ \delta }}}^H}{{\bf{\Phi }}^{c,H}}{{\bf{y}}^c}}}{{{{\big\| {{{\bf{\Phi }}^c}{\bm{ \delta }}} \big\|}^2}}}.
	\end{align}
	Substituting (\ref{Opti_gama}) into (\ref{P_jointML_1}) and ignoring irrelevant terms yields the following problem with respect to $ {{\bm{ \delta }}} $:
	\begin{subequations}\label{P_subproblem_delta}
		\begin{align}
		\mathop {\max }\limits_{{\bm{ \delta }}} & \ {\rm{ }}\frac{{{{{\bm{ \delta }}}^H}{\bf{\Xi }}_1^c{\bm{ \delta }}}}{{{{{\bm{ \delta }}}^H}{\bf{\Xi }}_2^c{\bm{ \delta }}}}\\
		{\rm{s}}{\rm{.t}}{\rm{. }} & \ \left( \text{\ref{P_jointML_1}b} \right),
		\end{align}
	\end{subequations}
	where $ {\bf{\Xi }}_1^c = {{\bf{\Phi }}^{c,H}}{{\bf{y}}^c}{{\bf{y}}^{c,H}}{{\bf{\Phi }}^c} $ and $ {\bf{\Xi }}_2^c = {{\bf{\Phi }}^{c,H}}{{\bf{\Phi }}^c} $.
	It can be readily seen that problem (\ref{P_subproblem_delta}) is a non-convex single-ratio fractional programming (FP) problem whose optimal solution is difficult to obtain. Moreover, we can see that the numerator of the fractional term in (\ref{P_subproblem_delta}a) is a nonnegative function, and the denominator is a positive function with probability one considering constraint (\ref{P_subproblem_delta}b). Therefore, to solve problem (\ref{P_subproblem_delta}), we apply the classical Dinkelbach's transform to decouple the numerator and denominator of the objective function (\ref{P_subproblem_delta}a) \cite{Dinkelbach,10218356_Wei}, thus leading to the following problem:
	\begin{subequations}\label{P_Dinkelbach}
		\begin{align}
		\mathop {\max }\limits_{{\bm{ \delta }}} & \  {\rm{ }}{{{\bm{ \delta }}}^H}\left( {{\bf{\Xi }}_1^c - y{\bf{\Xi }}_2^c} \right){\bm{ \delta }}\\
		{\rm{s}}{\rm{.t}}{\rm{. }}& \ \left( \text{\ref{P_jointML_1}b} \right),
		\end{align}
	\end{subequations}
	where $ y $ is an auxiliary variable, which is iteratively updated by
	\begin{align}
	{y^{\left( {t + 1} \right)}} = \frac{{{{{\bm{ \delta }}}^{\left( t \right),H}}{\bf{\Xi }}_1^c{{{\bm{ \delta }}}^{\left( t \right)}}}}{{{{{\bm{ \delta }}}^{\left( t \right),H}}{\bf{\Xi }}_2^c{{{\bm{ \delta }}}^{\left( t \right)}}}},
	\end{align}
	where $ t $ is the iteration index.
	The following proposition presents a sufficient condition for problem (\ref{P_subproblem_delta}) to obtain a globally optimal solution.
	\begin{theorem}
		When a globally optimal solution can be obtained by solving problem (\ref{P_Dinkelbach}) in each iteration, then the optimal solution of problem (\ref{P_Dinkelbach}) $ {{{\bm{ \delta }}}^\dag } $ corresponding to $ {y^\dag } $, which is a limit point of the sequence $ \big\{ {{y^{\left( t \right)}}} \big\} $, will be a globally optimal solution of problem (\ref{P_subproblem_delta}).
	\end{theorem}
	\begin{IEEEproof}
		We first show that the sequence $ \big\{ {{y^{\left( t \right)}}} \big\} $ is a monotonically convergent sequence.
		Since
		\begin{align}
		{{{\bm{ \delta }}}^{\left( {t + 1} \right)}} = \arg \mathop {\max }\limits_{{\bm{ \delta }}:{\bm{ \delta }}\left( n \right) \in \left\{ { - 1,1} \right\},\forall n} \big\{ {{{{\bm{ \delta }}}^H}\big( {{\bf{\Xi }}_1^c - {y^{\left( t \right)}}{\bf{\Xi }}_2^c} \big){\bm{ \delta }}} \big\},
		\end{align}
		then we have
		\begin{align}\label{Pro_2_2}
		{{{\bm{ \delta }}}^{\left( {t + 1} \right),H}}\big( {{\bf{\Xi }}_1^c - {y^{\left( t \right)}}{\bf{\Xi }}_2^c} \big){{{\bm{ \delta }}}^{\left( {t + 1} \right)}} \ge {{{\bm{ \delta }}}^{\left( t \right),H}}\big( {{\bf{\Xi }}_1^c - {y^{\left( t \right)}}{\bf{\Xi }}_2^c} \big){{{\bm{ \delta }}}^{\left( t \right)}} = 0.
		\end{align}
		By rearranging (\ref{Pro_2_2}), we can obtain
		\begin{align}
		{y^{\left( {t + 1} \right)}} = \frac{{{{{\bm{ \delta }}}^{\left( {t + 1} \right),H}}{\bf{\Xi }}_1^c{{{\bm{ \delta }}}^{\left( {t + 1} \right)}}}}{{{{{\bm{ \delta }}}^{\left( {t + 1} \right),H}}{\bf{\Xi }}_2^c{{{\bm{ \delta }}}^{\left( {t + 1} \right)}}}} \ge {y^{\left( t \right)}},
		\end{align}
		thus the sequence $ \big\{ {{y^{\left( t \right)}}} \big\} $ is monotonically non-decreasing. Moreover, 
		$ {{y^{\left( t \right)}}} $ is upper bounded. Therefore, the sequence $ \big\{ {{y^{\left( t \right)}}} \big\} $ is guaranteed to monotonically converge. 
		
		At any limit point $ \big\{ {{{{\bm{ \delta }}}^\dag },{y^\dag }} \big\} $, it is readily seen that $ {{{\bm{ \delta }}}^{\dag ,H}}\left( {{\bf{\Xi }}_1^c - {y^\dag }{\bf{\Xi }}_2^c} \right){{{\bm{ \delta }}}^\dag } = 0 $ holds, i.e., $ {y^\dag } = \frac{{{{{\bm{ \delta }}}^{\dag ,H}}{\bf{\Xi }}_1^c{{{\bm{ \delta }}}^\dag }}}{{{{{\bm{ \delta }}}^{\dag ,H}}{\bf{\Xi }}_2^c{{{\bm{ \delta }}}^\dag }}} $. We now prove the proposition by contradiction.  Suppose that $ {{{\bm{ \delta }}}^\dag } $ is not a globally optimal solution of problem (\ref{P_subproblem_delta}), then there must exist a $ {{{\bm{ \delta }}}^ * } $ such that $ \frac{{{{{\bm{ \delta }}}^{ * ,H}}{\bf{\Xi }}_1^c{{{\bm{ \delta }}}^ * }}}{{{{{\bm{ \delta }}}^{ * ,H}}{\bf{\Xi }}_2^c{{{\bm{ \delta }}}^ * }}} > \frac{{{{{\bm{ \delta }}}^{\dag ,H}}{\bf{\Xi }}_1^c{{{\bm{ \delta }}}^\dag }}}{{{{{\bm{ \delta }}}^{\dag ,H}}{\bf{\Xi }}_2^c{{{\bm{ \delta }}}^\dag }}} = {y^\dag } $, i.e., $ {{{\bm{ \delta }}}^{ * ,H}}\left( {{\bf{\Xi }}_1^c - {y^\dag }{\bf{\Xi }}_2^c} \right){{{\bm{ \delta }}}^ * } > 0 $, that is $ \mathop {\max }\limits_{{\bm{ \delta }}} \big\{ {{{{\bm{ \delta }}}^H}\left( {{\bf{\Xi }}_1^c - {y^\dag }{\bf{\Xi }}_2^c} \right){\bm{ \delta }}} \big\} > 0 $, which contradicts the fact that $ {{y^\dag }} $ is a limit point of $ \left\{ {{y^{\left( t \right)}}} \right\} $. This completes the proof.
	\end{IEEEproof}
	
	We can see that problem (\ref{P_Dinkelbach}) is a quadratic optimization problem with binary variables,
	which can be equivalently reformulated as an integer linear program (ILP) whose globally optimal solution can be obtained by applying the branch-and-bound method \cite{burer2012non,6119233_Joshi}.
	Let $ {{\bf{R}}^c} = {\bf{\Xi }}_1^c - y{\bf{\Xi }}_2^c $. By applying some algebraic manipulations using the fact that $ {{\bf{R}}^c} $ is a Hermitian matrix, the objective function of problem (\ref{P_Dinkelbach}) can be expanded as
	\begin{align}\label{obj_1}
	{{{\bm{ \delta }}}^H}{{\bf{R}}^c}{\bm{ \delta }} = \sum\limits_i {r_{ii}^c}  + \sum\limits_{i > j} {2\Re \left\{ {r_{ij}^c} \right\}{{ \delta }_i}{{ \delta }_j}},
	\end{align}
	where $ r_{ij}^c, \forall i,j \in {\cal N}, $ is the $ \left( {i,j} \right) $-th entry of $ {{\bf{R}}^c} $, and $ {{ \delta }_i} \in \left\{ { - 1,1} \right\},\forall i \in {\cal N}, $ is the $ i $-th entry of $ {{\bm{ \delta }}} $. Substituting $ {{ \delta }_i} = 2{\upsilon _i} - 1 $ into (\ref{obj_1}) where $ {\upsilon _i} \in \left\{ {0,1} \right\} $, the objective function of problem (\ref{P_Dinkelbach}) can be equivalently rewritten as
	(by ignoring the constant terms)
	\begin{align}\label{obj_2}
	{{\cal G}_1}\left( {\left\{ {{\upsilon _i}} \right\}} \right) = 8\sum\limits_{i > j} {\Re \left\{ {r_{ij}^c} \right\}{\upsilon _i}{\upsilon _j}}  - 4\sum\limits_{i > j} {\Re \left\{ {r_{ij}^c} \right\}\left( {{\upsilon _i} + {\upsilon _j}} \right)}.
	\end{align}
	By replacing the term of the quadratic form $ {\upsilon _i}{\upsilon _j},\forall i \ne j, $ in (\ref{obj_2}) with a new binary variable $ {\upsilon _{ij}} $, along with constraints $ {\upsilon _{ij}} \le {\upsilon _i} $, $ {\upsilon _{ij}} \le {\upsilon _j} $, $ {\upsilon _{ij}} \ge {\upsilon _i} + {\upsilon _j} - 1 $ and $ {\upsilon _{ij}} \in \left\{ {0,1} \right\} $, problem (\ref{P_Dinkelbach}) can be finally reformulated as
	\begin{subequations}\label{P_ILP}
		\begin{align}
		\mathop {\max }\limits_{\left\{ {{\upsilon _{ij}}} \right\},\left\{ {{\upsilon _i}} \right\}} {\rm{ }} & \ 8\sum\limits_{i > j} {\Re \left\{ {r_{ij}^c} \right\}{\upsilon _{ij}}}  - 4\sum\limits_{i > j} {\Re \left\{ {r_{ij}^c} \right\}\left( {{\upsilon _i} + {\upsilon _j}} \right)} \\
		{\rm{s}}{\rm{.t}}{\rm{. }} & \ {\upsilon _{ij}} \ge {\upsilon _i} + {\upsilon _j} - 1, \\
		& \ {\upsilon _{ij}} \le {\upsilon _i},\\
		& \ {\upsilon _{ij}} \le {\upsilon _j},\\
		& \ {\upsilon _i} \in \left\{ {0,1} \right\}, \forall i \in {\cal N},\\
		& \ {\upsilon _{ij}} \in \left\{ {0,1} \right\},\forall i,j \in {\cal N},i \ne j.
		\end{align}
	\end{subequations}
	It can be readily seen that problem (\ref{P_ILP}) is an ILP and thus can be optimally solved by applying the branch-and-bound technique. 
	
	With the obtained $ {{{\hat \gamma }^c}} $ and $ {{{{\bm{\hat \delta }}}^c}} $, the estimated value of $ \alpha $ under the hypothesis $ j_T $ in the $ c $-th cycle can be calculated as
	\begin{align}
	{{\hat \alpha }^c} = \frac{{{{\hat \gamma }^c}}}{{{{\bf{a}}^T}\left( {{j_T}} \right){{\bf{\Theta }}^c}{\rm{diag}}\big( {{{{\bm{\hat \delta }}}^c}} \big){\bf{\hat G}}{{\bf{x}}^c}}}.
	\end{align}
	
	\section{Joint Optimization of Transmit Waveform and IRS Phase Shifts}
	In this section, we formulate the joint BS transmit waveform and IRS phase shifts optimization problem, which helps to determine the transmit signals of the BS and IRS phase shifts for the next target localization cycle based on the target localization results of the current cycle, thus improving the performance of the proposed target localization scheme.
	
	\subsection{Problem Formulation}
	In our work, we resort to relative entropy, which is an information-theoretic metric, to indicate the ``distance" between the probability functions of two different hypotheses \cite{9725255_Zhang,8101018_Tang}. The maximization of the ``distance" allows us to distinguish more easily between the two hypotheses, and thus we can obtain a higher localization accuracy. Similar to \cite{9725255_Zhang}, the distance between hypotheses $ i_T $ and $ j_T $ in the $ \left( {c + 1} \right) $-th cycle of the localization is defined as
	\begin{align}
	&d_{{i_T},{j_T}}^{c + 1}\left( {{{\cal P}^{c + 1}}} \right) \notag\\
	& = D\left( {{p^{c + 1}}\left( {{\bf{y}}|{i_T},{{\cal P}^{c + 1}}} \right)\parallel {p^{c + 1}}\left( {{\bf{y}}|{j_T},{{\cal P}^{c + 1}}} \right)} \right) \notag \\
	& \quad + D\left( {{p^{c + 1}}\left( {{\bf{y}}|{j_T},{{\cal P}^{c + 1}}} \right)\parallel {p^{c + 1}}\left( {{\bf{y}}|{i_T},{{\cal P}^{c + 1}}} \right)} \right),
	\end{align}
	where $ {{\cal P}^{c + 1}} = \left\{ {{{\bf{x}}^{c + 1}},{{\bm{\theta }}^{c + 1}}} \right\} $ is the set of variables to be optimized in the $ \left( {c + 1} \right) $-th cycle, and $ {{p^{c + 1}}\left( {{\bf{y}}|{i_T},{{\cal P}^{c + 1}}} \right)} $ is the probability function given hypothesis $ i_T $, variables $ {{\cal P}^{c + 1}} $, estimated $ {\hat \alpha } $ and $ {{\bm{\hat \delta }}} $ in the $ c $-th cycle. Moreover, $ D\left( {{p^{c + 1}}\left( {{\bf{y}}|{i_T},{{\cal P}^{c + 1}}} \right)\parallel {p^{c + 1}}\left( {{\bf{y}}|{j_T},{{\cal P}^{c + 1}}} \right)} \right) $ is the relative entropy between probability functions $ {{p^{c + 1}}\left( {{\bf{y}}|{i_T},{{\cal P}^{c + 1}}} \right)} $ and $ {{p^{c + 1}}\left( {{\bf{y}}|{j_T},{{\cal P}^{c + 1}}} \right)} $, i.e.,
	\begin{align}
	&D\left( {{p^{c + 1}}\left( {{\bf{y}}|{i_T},{{\cal P}^{c + 1}}} \right)\parallel {p^{c + 1}}\left( {{\bf{y}}|{j_T},{{\cal P}^{c + 1}}} \right)} \right) \notag\\
	&= \int {{p^{c + 1}}\left( {{\bf{y}}|{i_T},{{\cal P}^{c + 1}}} \right)\log \frac{{{p^{c + 1}}\left( {{\bf{y}}|{i_T},{{\cal P}^{c + 1}}} \right)}}{{{p^{c + 1}}\left( {{\bf{y}}|{j_T},{{\cal P}^{c + 1}}} \right)}}} d{\bf{y}}.
	\end{align}
	According to \cite{9725255_Zhang}, we can simplify the expression for $ d_{{i_T},{j_T}}^{c + 1}\left( {{{\cal P}^{c + 1}}} \right) $ as
	\begin{align}
	d_{{i_T},{j_T}}^{c + 1}\left( {{{\cal P}^{c + 1}}} \right) = \frac{1}{{{\sigma ^2}}}{\left\| {{\bf{\bar y}}\left( {{i_T},{{\cal P}^{c + 1}}} \right) - {\bf{\bar y}}\left( {{j_T},{{\cal P}^{c + 1}}} \right)} \right\|^2},
	\end{align}
	where
	\begin{align}
	&{\bf{\bar y}}\left( {{i_T},{{\cal P}^{c + 1}}} \right) \notag\\
	&= {\rm{vec}}\big( {\hat \alpha _{{i_T}}^c{{\bf{\hat G}}_{{i_T}}^{c,T}}{{\bf{\Theta }}^{c + 1}}{{\bf{a}}_{{i_T}}}{\bf{a}}_{{i_T}}^T{{\bf{\Theta }}^{c + 1}}{{\bf{\hat G}}_{{i_T}}^c}{{\bf{X}}^{c + 1}}} \big),\forall i \in {\cal I},
	\end{align}
	where we define $ {\bf{\hat G}}_{{i_T}}^c \buildrel \Delta \over = {\rm{diag}}\big( {{{{\bm{\hat \delta }}}^c}\left( {{i_T}} \right)} \big){\bf{\hat G}} $, and we simplify the terms $ {{\hat \alpha }^c}\left( {{i_T}} \right) $ and $ {\bf{a}}\left( {{i_T}} \right) $ into $ \hat \alpha _{{i_T}}^c $ and $ {{{\bf{a}}_{{i_T}}}} $, respectively.
	Our objective is to maximize the weighted sum distance between each two hypotheses. Thus, the optimization problem for determining the BS transmit waveform and IRS phase shifts in the $ \left( {c + 1} \right) $-th cycle can be formulated as
	\begin{subequations}\label{P0}
		\begin{align}
		\mathop {\max }\limits_{{{\bf{x}}^{c + 1}},{{\bm{\theta }}^{c + 1}}} & \ \sum\limits_{i = 1}^I {\sum\limits_{j = i + 1}^I {\beta _{{i_T},{j_T}}^{c + 1}d_{{i_T},{j_T}}^{c + 1}} } \\
		{\rm{s}}{\rm{.t}}{\rm{. }} & \ {\left\| {{{\bf{x}}^{c + 1}}} \right\|^2} \le {P_b},\\
		& \ \left| {{{\bm{\theta }}^{c + 1}}\left( n \right)} \right| = 1,\forall n \in {\cal N},
		\end{align}
	\end{subequations}
	where constraint (\ref{P0}b) is the power constraint for the BS, with $ P_b $ being its allowed maximum transmit power. $ {\beta _{{i_T},{j_T}}^{c + 1}} $ is the weight/priority of the distance between hypotheses $ i_T $ and $ j_T $, which is set as the prior probability product of the two hypotheses as in \cite{9725255_Zhang}, i.e.,
	\begin{align}
	\beta _{{i_T},{j_T}}^{c + 1} = {p^{c + 1}}\left( {{i_T}} \right){p^{c + 1}}\left( {{j_T}} \right).
	\end{align}
	
	\emph{Remark 2:}
	It is challenging to solve problem (\ref{P0}) since $ d_{{i_T},{j_T}}^{c + 1}\left( {{{\cal P}^{c + 1}}} \right) $ is essentially a quartic function of $ {{\bm{\theta }}^{c + 1}} $. 
	In contrast, the IRS reflection coefficient matrices corresponding to the transmission and reception steps in \cite{9725255_Zhang} are designed separately, resulting in the objective function of the formulated optimization problem being a quadratic function of the IRS phase shifts, which is much easier to solve.
	
	\subsection{Problem Transformation \& Solution}
	Prior to solving problem (\ref{P0}), we present a more tractable expression for $ d_{{i_T},{j_T}}^{c + 1}\left( {{{\cal P}^{c + 1}}} \right) $ in the following Proposition \ref{func_phi}.
	
	\begin{theorem}\label{func_phi}
		The distance $ d_{{i_T},{j_T}}^{c + 1}\left( {{{\cal P}^{c + 1}}} \right) $ between hypotheses $ i_T $ and $ j_T $ can be expressed as
		\begin{align}
		\varphi _{{i_T},{j_T}}^{c + 1} = & \frac{{{L}}}{{{\sigma ^2}}}\left( {{{\left| {\hat \alpha _{{i_T}}^c} \right|}^2}{\rm{tr}}\left( {{{\bf{Q}}^{c + 1,H}}{{\bf{A}}_{{i_T},{i_T}}}{{\bf{Q}}^{c + 1}}{{\bf{B}}_{{i_T},{i_T}}}} \right)} \right. \notag\\
		& - 2\Re \left\{ {\hat \alpha _{{i_T}}^c\hat \alpha _{{j_T}}^{c, * }{\rm{tr}}\left( {{{\bf{Q}}^{c + 1,H}}{{\bf{A}}_{{i_T},{j_T}}}{{\bf{Q}}^{c + 1}}{{\bf{B}}_{{i_T},{j_T}}}} \right)} \right\} \notag\\
		& \left. { + {{\left| {\hat \alpha _{{j_T}}^c} \right|}^2}{\rm{tr}}\left( {{{\bf{Q}}^{c + 1,H}}{{\bf{A}}_{{j_T},{j_T}}}{{\bf{Q}}^{c + 1}}{{\bf{B}}_{{j_T},{j_T}}}} \right)} \right),
		\end{align}
		where $ {{\bf{Q}}^{c + 1}} = {{\bm{\theta }}^{c + 1}}{{\bm{\theta }}^{c + 1,H}} $, $ {{\bf{A}}_{{i_T},{j_T}}} = \big( {{\bf{\hat G}}_{{j_T}}^{c, * }{\bf{\hat G}}_{{i_T}}^{c,T}} \big) \odot {\left( {{{\bf{a}}_{{i_T}}}{\bf{a}}_{{j_T}}^H} \right)^T} $ and $ {{\bf{B}}_{{i_T},{j_T}}} = \left( {{\bf{a}}_{{j_T}}^ * {\bf{a}}_{{i_T}}^T} \right) \odot {\big( {{\bf{\hat G}}_{{i_T}}^c{{\bf{x}}^{c + 1}}{{\bf{x}}^{c + 1,H}}{\bf{\hat G}}_{{j_T}}^{c,H}} \big)^T}, \forall i,j \in \mathcal{I} $.
	\end{theorem}
	
	\begin{IEEEproof}
		We omit the proof for brevity.
	\end{IEEEproof}
	
	Based on Proposition \ref{func_phi}, problem (\ref{P0}) can be equivalently reformulated as
	\begin{subequations}\label{P1}
		\begin{align}
		\mathop {\max }\limits_{{{\bf{Q}}^{c + 1}},{{\bm{\theta }}^{c + 1}},{{\bf{x}}^{c + 1}}} & \  {\rm{ }}\sum\limits_{i = 1}^I {\sum\limits_{j = i + 1}^I {\beta _{{i_T},{j_T}}^{c + 1}\varphi _{{i_T},{j_T}}^{c + 1}} } \\
		{\rm{s}}{\rm{.t}}{\rm{. }} & \ {{\bf{Q}}^{c + 1}} = {{\bm{\theta }}^{c + 1}}{{\bm{\theta }}^{c + 1,H}},\\
		& \ \text{(\ref{P0}b)},\text{(\ref{P0}c)}.
		\end{align}
	\end{subequations}
	According to Lemma 1 in \cite{10274990_Ji}, constraint (\ref{P1}b) can be rewritten in an equivalent form as
	\begin{align}\label{constraint_Q}
	\Re \left\{ {{{\bm{\theta }}^{c + 1,H}}{{\bf{Q}}^{c + 1}}{{\bm{\theta }}^{c + 1}}} \right\} - {N^2} \ge 0,{{\bf{Q}}^{c + 1}} \in {\cal M},
	\end{align}
	where $ {{\bf{Q}}^{c + 1}} \in {\cal M} $ denotes that each entry of $ {{\bf{Q}}^{c + 1}} $ has a unit modulus constraint, i.e., $ \left| {{{\bf{Q}}^{c + 1}}\left( {i,j} \right)} \right| = 1,\forall i,j \in {\cal N} $. We apply a penalty-based method as in \cite{NO_Nocedal} to deal with the non-convex constraint (\ref{constraint_Q}), where the optimization variables $ {{{\bf{Q}}^{c + 1}}} $ and $ {{{\bm{\theta }}^{c + 1}}} $ are coupled in it. Specifically, by adding the constraint $ \Re \left\{ {{{\bm{\theta }}^{c + 1,H}}{{\bf{Q}}^{c + 1}}{{\bm{\theta }}^{c + 1}}} \right\} - {N^2} \ge 0 $ as a penalty term in the objective function of problem (\ref{P1}), we can obtain the following problem:
	\begin{subequations}\label{P2}
		\begin{align}
		\mathop {\max }\limits_{{{\bf{Q}}^{c + 1}},{{\bm{\theta }}^{c + 1}},{{\bf{x}}^{c + 1}}} & \  {\rm{ }}\sum\limits_{i = 1}^I {\sum\limits_{j = i + 1}^I {\beta _{{i_T},{j_T}}^{c + 1}\varphi _{{i_T},{j_T}}^{c + 1}} } \notag\\
		& \ \ + \frac{1}{{2\rho }}\left( {\Re \left\{ {{{\bm{\theta }}^{c + 1,H}}{{\bf{Q}}^{c + 1}}{{\bm{\theta }}^{c + 1}}} \right\} - {N^2}} \right)\\
		{\rm{s}}{\rm{.t}}{\rm{. }} & \ {{\bf{Q}}^{c + 1}} \in {\cal M},\\
		& \ \text{(\ref{P0}b)},\text{(\ref{P0}c)},
		\end{align}
	\end{subequations}
	where $ \rho > 0 $ is the penalty parameter.
	In the following, we propose an algorithm to obtain a high-quality solution to problem (\ref{P2}), which consists of a double-layer structure. Specifically, in the inner layer, problem (\ref{P2}) with fixed $ \rho $ is solved. While in the outer layer, we gradually decrease the value of $ \rho $, until the solution obtained by solving problem (\ref{P2}) satisfies the equality constraint (\ref{P1}b). The details are given as follows.
	
	In the inner layer, with any given $ \rho $, problem (\ref{P2}) is still a non-convex problem due to the non-convex objective function and non-convex constraints in (\ref{P2}b) and (\ref{P0}c). It can be observed that all the constraints in problem (\ref{P2}) are separable. Hence, we can apply the block coordination descent (BCD) method to address problem (\ref{P2}) with the block variables $ {{{\bf{Q}}^{c + 1}}} $, $ {{{\bf{x}}^{c + 1}}} $, and $ {{{\bm{\theta }}^{c + 1}}} $, i.e., we optimize one block variable among them while fixing the others each time, and iterate the procedure until the convergence is reached. Therefore, we need to solve the following three sub-problems in each BCD iteration:
	\subsubsection{The Sub-problem w.r.t $ {{{\bf{Q}}^{c + 1}}} $}
	For the given blocks $ {{{\bf{x}}^{c + 1}}} $ and $ {{{\bm{\theta }}^{c + 1}}} $, the variable $ {{{\bf{Q}}^{c + 1}}} $ is updated by solving the following problem (with constant terms ignored)
	\begin{align}
	\mathop {\max }\limits_{{{\bf{Q}}^{c + 1}} \in {\cal M}} {\rm{ }} \varpi \left( {{{\bf{Q}}^{c + 1}}} \right),
	\end{align}
	where $ \varpi \left( {{{\bf{Q}}^{c + 1}}} \right) = \sum\limits_{i = 1}^I {\sum\limits_{j = i + 1}^I {\beta _{{i_T},{j_T}}^{c + 1}\varphi _{{i_T},{j_T}}^{c + 1}} }  + \frac{1}{{2\rho }}\Re \left\{ {{{\bm{\theta }}^{c + 1,H}}{{\bf{Q}}^{c + 1}}{{\bm{\theta }}^{c + 1}}} \right\} $. We also apply the BCD method to optimize $ {{{\bf{Q}}^{c + 1}}} $, i.e., we update one entry of $ {{{\bf{Q}}^{c + 1}}} $ each time while fixing the others, until $ {{{\bf{Q}}^{c + 1}}} $ converges. The following proposition gives how the entries in $ {{{\bf{Q}}^{c + 1}}} $ are updated.
	\begin{theorem}
		When the BCD method is applied to optimize $ {{{\bf{Q}}^{c + 1}}} $, one element of $ {{{\bf{Q}}^{c + 1}}} $ will be updated at each iteration according to (\ref{update_Q}), which is shown at the top of the next page,
		\begin{figure*}[!t]
			\normalsize
			\begin{align}\label{update_Q}
			{{\bf{Q}}^{c + 1}}\left( {m,n} \right) \leftarrow & \exp \left\{ {j\arg \left( {\sum\limits_{j > i} {\beta _{{i_T},{j_T}}^{c + 1}{{\left[ {\frac{{\partial \varphi _{{i_T},{j_T}}^{c + 1}}}{{\partial {{\bf{Q}}^{c + 1, * }}}}} \right]}_{m,n}}}  + \frac{1}{{4\rho }}{{\left[ {{{\bm{\theta }}^{c + 1}}{{\bm{\theta }}^{c + 1,H}}} \right]}_{m,n}} - \frac{{{L}}}{{{\sigma ^2}}}{{\bf{Q}}^{c + 1}}\left( {m,n} \right) \sum\limits_{j > i} {\beta _{{i_T},{j_T}}^{c + 1}\chi _{{i_T},{j_T}}^c} } \right)} \right\}, \notag\\
			&\quad \qquad \qquad \qquad \qquad \qquad \qquad \qquad \qquad \qquad \qquad \qquad \qquad \qquad \qquad \qquad \qquad \qquad \qquad \forall m,n \in {\cal N}.
			\end{align}
			
			\hrulefill
			\vspace*{-12pt}
			
			\begin{align}\label{partial_derivative}
			\frac{{\partial \varphi _{{i_T},{j_T}}^{c + 1}}}{{\partial {{\bf{Q}}^{c + 1, * }}}} =& \frac{{{L}}}{{{\sigma ^2}}}\left( {{{\left| {\hat \alpha _{{i_T}}^c} \right|}^2}{{\bf{A}}_{{i_T},{i_T}}}{{\bf{Q}}^{c + 1}}{{\bf{B}}_{{i_T},{i_T}}} - \hat \alpha _{{i_T}}^c\hat \alpha _{{j_T}}^{c, * }{{\bf{A}}_{{i_T},{j_T}}}{{\bf{Q}}^{c + 1}}{{\bf{B}}_{{i_T},{j_T}}}} \right. \notag\\
			&\qquad \qquad \qquad \qquad \qquad \qquad \qquad \qquad \qquad \left. { - \hat \alpha _{{i_T}}^{c, * }\hat \alpha _{{j_T}}^c{\bf{A}}_{{i_T},{j_T}}^H{{\bf{Q}}^{c + 1}}{\bf{B}}_{{i_T},{j_T}}^H + {{\left| {\hat \alpha _{{j_T}}^c} \right|}^2}{{\bf{A}}_{{j_T},{j_T}}}{{\bf{Q}}^{c + 1}}{{\bf{B}}_{{j_T},{j_T}}}} \right). 
			\end{align}
			
			\hrulefill
			\vspace*{-12pt}
			
			\begin{align}\label{kai}
			\chi _{{i_T},{j_T}}^c = {\left| {\hat \alpha _{{i_T}}^c} \right|^2}{{\bf{A}}_{{i_T},{i_T}}}\left( {m,m} \right){{\bf{B}}_{{i_T},{i_T}}}\left( {n,n} \right) - 2\Re \left\{ {\hat \alpha _{{i_T}}^c\hat \alpha _{{j_T}}^{c, * }{{\bf{A}}_{{i_T},{j_T}}}\left( {m,m} \right){{\bf{B}}_{{i_T},{j_T}}}\left( {n,n} \right)} \right\} + {\left| {\hat \alpha _{{j_T}}^c} \right|^2}{{\bf{A}}_{{j_T},{j_T}}}\left( {m,m} \right){{\bf{B}}_{{j_T},{j_T}}}\left( {n,n} \right).
			\end{align}
			\hrulefill
		\end{figure*}
		where $ \frac{{\partial \varphi _{{i_T},{j_T}}^{c + 1}}}{{\partial {{\bf{Q}}^{c + 1, * }}}} $ and $ {\chi _{{i_T},{j_T}}^c} $ are shown in (\ref{partial_derivative}) and (\ref{kai}), respectively.
	\end{theorem}
	
	\begin{IEEEproof}
		The derivation can be performed as in \cite{10274990_Ji}, which is omitted here for brevity.
	\end{IEEEproof}
	
	\subsubsection{The Sub-problem w.r.t $ {{\bf{x}}^{c + 1}} $}
	Since $ {{{\bf{B}}_{{i_T},{j_T}}}}, \forall i,j \in \mathcal{I} $, is a function of $ {{\bf{x}}^{c + 1}} $, we first reformulate the term $ {\rm{tr}}\left( {{{\bf{Q}}^{c + 1,H}}{{\bf{A}}_{{i_T},{j_T}}}{{\bf{Q}}^{c + 1}}{{\bf{B}}_{{i_T},{j_T}}}} \right) $ in $ \varphi _{{i_T},{j_T}}^{c + 1} $ as
	\begin{align}
	{\rm{tr}}\left( {{{\bf{Q}}^{c + 1,H}}{{\bf{A}}_{{i_T},{j_T}}}{{\bf{Q}}^{c + 1}}{{\bf{B}}_{{i_T},{j_T}}}} \right) = {{\bf{x}}^{c + 1,H}}{\bf{\Psi }}_{{i_T},{j_T}}^{c + 1}{{\bf{x}}^{c + 1}},
	\end{align}
	where
	\begin{align}
	&{\bf{\Psi }}_{{i_T},{j_T}}^{c + 1} = {\bf{\hat G}}_{{j_T}}^{c,H}\left( {\left( {{{\bf{Q}}^{c + 1,T}}{\bf{A}}_{{i_T},{j_T}}^T{{\bf{Q}}^{c + 1, * }}} \right) \odot \left( {{\bf{a}}_{{j_T}}^ * {\bf{a}}_{{i_T}}^T} \right)} \right){\bf{\hat G}}_{{i_T}}^c, \notag\\
	& \qquad \qquad \qquad \qquad \qquad \qquad \qquad \qquad \qquad  \forall i,j \in {\cal I}.
	\end{align}
	The derivation is omitted for brevity.
	Therefore, for the given blocks $ {{{\bf{Q}}^{c + 1}}} $ and $ {{{\bm{\theta }}^{c + 1}}} $, the variable $ {{\bf{x}}^{c + 1}} $ is updated by solving the following problem (with constant terms ignored)
	\begin{subequations}\label{subproblem_x}
		\begin{align}
		\mathop {\max }\limits_{{{\bf{x}}^{c + 1}}} & \ {\rm{ }}{{\bf{x}}^{c + 1,H}}{{\bf{Z}}^{c + 1}}{{\bf{x}}^{c + 1}}\\
		{\rm{s}}{\rm{.t}}{\rm{. }} & \ \text{(\ref{P0}b)},
		\end{align}
	\end{subequations}
	where
	\begin{align}
	{{\bf{Z}}^{c + 1}} = \sum\limits_{j > i} {\beta _{{i_T},{j_T}}^{c + 1}\big( {{{\left| {\hat \alpha _{{i_T}}^c} \right|}^2}{\bf{\Psi }}_{{i_T},{i_T}}^{c + 1} - \hat \alpha _{{i_T}}^c\hat \alpha _{{j_T}}^{c, * }{\bf{\Psi }}_{{i_T},{j_T}}^{c + 1}} \big.} \quad & \notag\\
	\big. { - \hat \alpha _{{i_T}}^{c, * }\hat \alpha _{{j_T}}^c{\bf{\Psi }}_{{j_T},{i_T}}^{c + 1} + {{\left| {\hat \alpha _{{j_T}}^c} \right|}^2}{\bf{\Psi }}_{{j_T},{j_T}}^{c + 1}} \big).&
	\end{align}
	It can be readily seen that the optimal solution to problem (\ref{subproblem_x}) is given by 
	\begin{align}
	{{\bf{x}}^{c + 1,{\rm{opt}}}} = \sqrt {{P_b}} {{\bm{\lambda }}_{\max }}\left( {{{\bf{Z}}^{c + 1}}} \right),
	\end{align}
	where $ {{\bm{\lambda }}_{\max }}\left( {{{\bf{Z}}^{c + 1}}} \right) $ is the eigenvector
	corresponding to the dominant eigenvalue of $ {{{\bf{Z}}^{c + 1}}} $.
	
	\subsubsection{The Sub-problem w.r.t $ {{{\bm{\theta }}^{c + 1}}} $}
	For given blocks $ {{{\bf{Q}}^{c + 1}}} $ and $ {{\bf{x}}^{c + 1}} $, the optimization variable $ {{{\bm{\theta }}^{c + 1}}} $ can be updated by solving the following problem
	\begin{subequations}\label{subproblem_theta}
		\begin{align}
		\mathop {\max }\limits_{{{\bm{\theta }}^{c + 1}}} & \ {\rm{ }}{{\bm{\theta }}^{c + 1,H}}{\bf{{P}}}{{\bm{\theta }}^{c + 1}}\\
		{\rm{s}}{\rm{.t}}{\rm{.}} & \ \text{(\ref{P0}c)},
		\end{align}
	\end{subequations}
	where $ {\bf{P}} = \left( {{{\bf{Q}}^{c + 1,H}} + {{\bf{Q}}^{c + 1}}} \right)/2 $. We can also apply the BCD algorithm to solve problem (\ref{subproblem_theta}). The following proposition shows how the entries in $ {{{\bm{\theta }}^{c + 1}}} $ are updated.
	\begin{theorem}
		When the BCD method is applied to optimize $ {{{\bm{\theta }}^{c + 1}}} $, one element of $ {{{\bm{\theta }}^{c + 1}}} $ will be updated at each iteration as
		\begin{align}
		&{{\bm{\theta }}^{c + 1}}\left( m \right) \leftarrow \exp \left\{ {j\arg \left( {{\bf{P}}\left( {m,:} \right){{\bm{\theta }}^{c + 1}}} \right.} \right. \notag\\
		& \qquad \qquad \qquad \quad \left. {\left. { - {\bf{P}}\left( {m,m} \right){{\bm{\theta }}^{c + 1}}\left( m \right)} \right)} \right\},\forall m \in {\cal N}.
		\end{align}
	\end{theorem}
	
	\begin{IEEEproof}
		The derivation is similar to that of $ {{{\bf{Q}}^{c + 1}}} $, we omit it here for brevity.
	\end{IEEEproof}
	
	In the outer layer, we gradually decrease the value of the penalty parameter $ \rho $ as
	\begin{align}
	\rho  \leftarrow c\rho ,0 < c < 1,
	\end{align}
	where $ c $ is a constant scaling factor. We also define the constraint violation indicator as
	\begin{align}
	\xi  = \frac{{{N^2} - \Re \left\{ {{{\bm{\theta }}^{c + 1,H}}{{\bf{Q}}^{c + 1}}{{\bm{\theta }}^{c + 1}}} \right\}}}{{{N^2}}}.
	\end{align}
	Our proposed penalty-based algorithm will be terminated when the inequality $ \xi  < \epsilon  $ is satisfied, where $ \epsilon $ is a predefined accuracy value.
	
	\begin{figure}[H]
		\centerline{\includegraphics[width=0.48\textwidth]{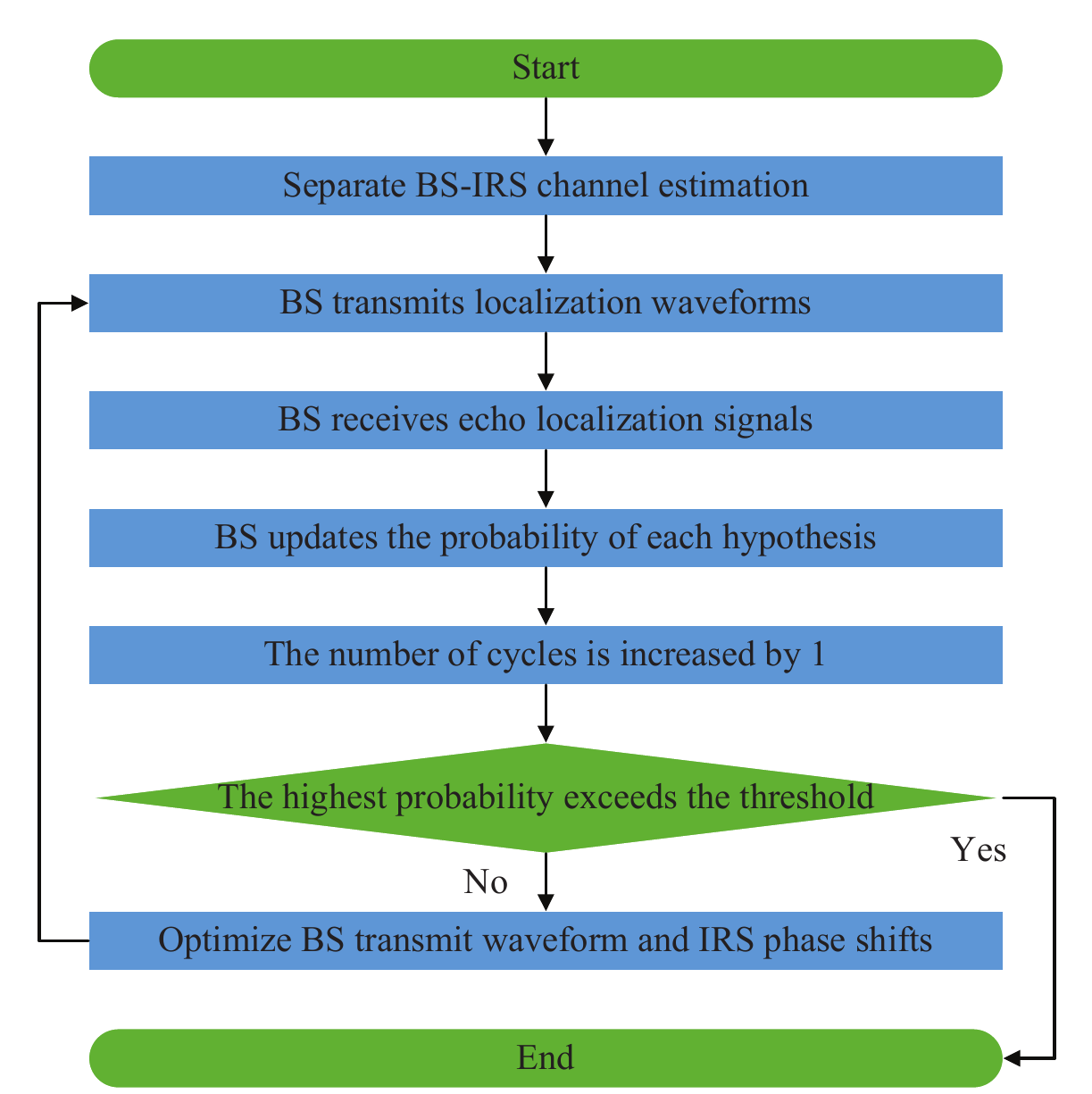}}
		\caption{Flow chart of the proposed overall target localization procedure.}
		\label{fig_FlowChart}
	\end{figure}
	
	The proposed overall target localization procedure is summarized in Fig. \ref{fig_FlowChart}. The localization procedure terminates when the highest hypothesis probability exceeds a predetermined threshold, for example $ 95\% $, and the hypothesis with the highest probability is selected as the final localization result.
	
	\begin{figure}[H]
		\centerline{\includegraphics[width=0.45\textwidth]{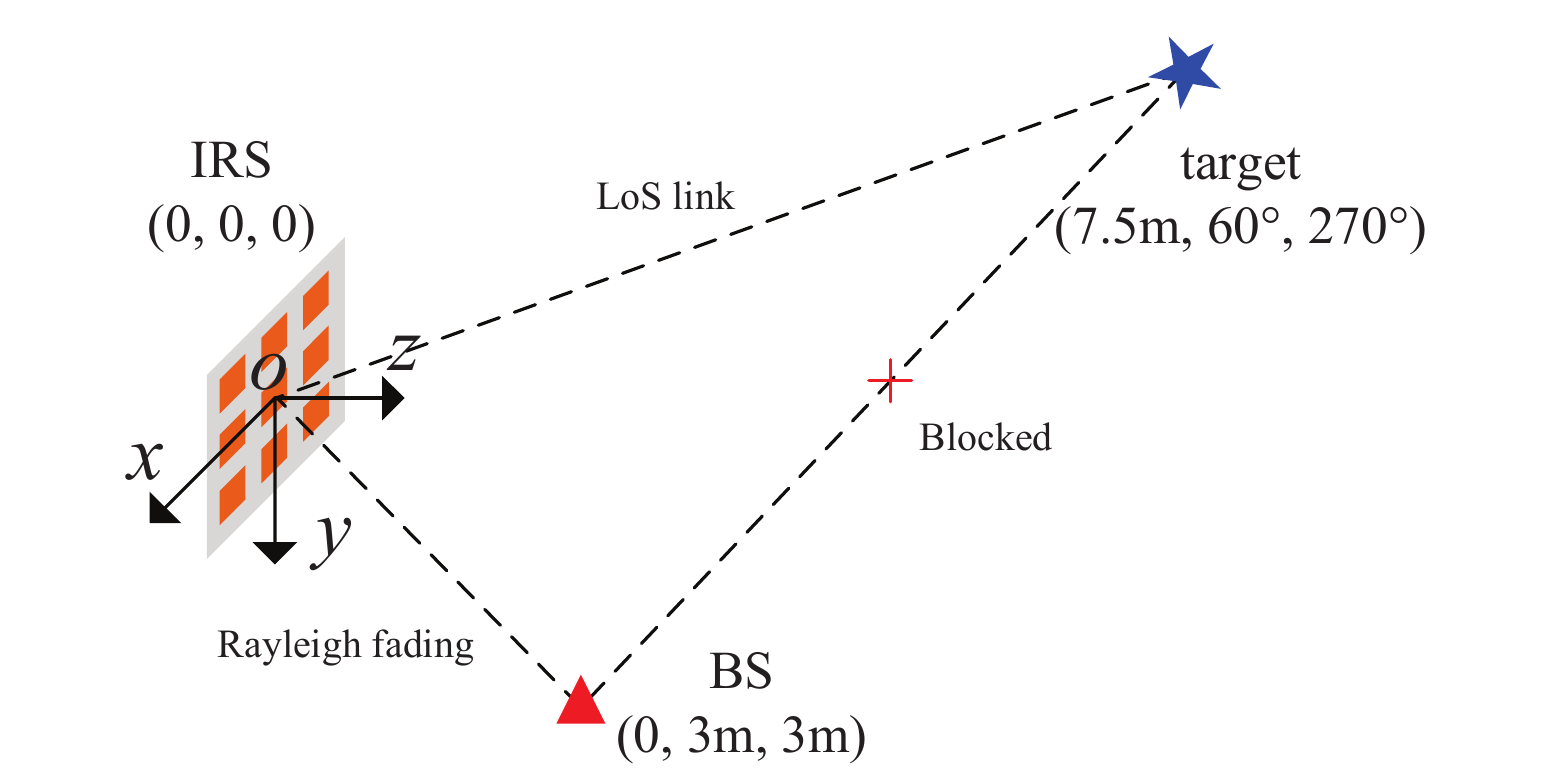}}
		\caption{Simulation setup.}
		\label{fig_simulation}
	\end{figure}
	
	\section{Numerical Results}
	
	\subsection{Simulation setup}
	In this section, simulation results are provided to evaluate the effectiveness of our proposed IRS-aided target localization scheme. The configuration of the system is shown in Fig.~\ref{fig_simulation}. Specifically, the IRS lies in the $ xoy $-plane with its center located at $ \left( {0,0,0} \right) $, the BS is located at $ \left( {0,3{\text{ m}}, 3{\text{ m}}} \right) $, and the target is located at $ \left( {7.5{\text{ m}},{{60}^ \circ },{{270}^ \circ }} \right) $.
	For the BS-IRS and IRS-target links, the distance-dependent path loss is modeled as $ L\left( d \right) = {C_0}{\left( d/d_0 \right)^{ - \alpha_0 }} $,
	where $ d $ is the individual transmission link distance, $ C_0 = -30 \text{ dB} $ denotes the channel power gain at the reference distance $ d_0 = 1 \text{ m} $, and the path-loss exponent $ \alpha_0 $ is set to $ 2.2 $. For simplicity, the small-scale fading of the BS-IRS channel follows Rayleigh fading, and the LoS channel dominates the IRS-target link. In addition, we assume that $ {\rm{vec}}\left( {{\bf{H}}_{{\cal A},{\cal B}}^{{\rm{SI}}}} \right) \sim {\cal C}{\cal N}\left( {{\bf{0}},\sigma _{{\rm{SI}}}^2{\bf{I}}} \right) $ and $ {\rm{vec}}\left( {{\bf{H}}_{{\cal A},{\cal B}}^{{\rm{Ref}}}} \right) \sim {\cal C}{\cal N}\left( {{\bf{0}},\sigma _{{\rm{Ref}}}^2{\bf{I}}} \right) $, and we set $ \sigma _{{\rm{Ref}}}^2 = \sigma _{{\rm{SI}}}^2 =  - 10{\text{ dB}} $ \cite{8931772_Sharma} without loss of generality.
	The amplitude of the target response is set to $ 1 $, and the phase of the target response is randomly selected in $ \left[ {0,2\pi } \right) $ in each Monte Carlo run. The angular range of interest is $ \theta  \in \left[ {{{52.5}^ \circ },{{72.5}^ \circ }} \right),\phi  = {270^ \circ } $, and the range $ \left[ {{{52.5}^ \circ },{{72.5}^ \circ }} \right) $ is uniformly divided into $ I = 4 $ grids, namely $ \left[ {{{52.5}^ \circ },{{57.5}^ \circ }} \right) $, $ \left[ {{{57.5}^ \circ },{{62.5}^ \circ }} \right) $, $ \left[ {{{62.5}^ \circ },{{67.5}^ \circ }} \right) $, and $ \left[ {{{67.5}^ \circ },{{72.5}^ \circ }} \right) $.
	We denote the hypotheses that the target is located at the above four grids as $ H_1 $, $ H_2 $, $ H_3 $, and $ H_4 $, respectively. Obviously, the correct hypothesis is $ H_2 $.
	Moreover, the number of snapshots of the BS transmit waveform is set to $ L = 8 $.
	Unless otherwise specified, we fix $ N_x = 5 $ and increase $ N_y $ linearly with $ N $, and the number of BS antennas and noise power are set to $ M = 4 $ and $ {\sigma ^2} =  - 120{\text{ dBm}} $ \cite{10138058_Song}, respectively. All the results are averaged over 30 independent BS-IRS channel realizations.
	
	\subsection{BS-IRS Channel Estimation}
	We define the received signal-to-noise ratio (SNR) by considering the BS-IRS round-trip channel fading as
	\begin{align}
	{{\mathop{\rm SNR}\nolimits} _r} = \frac{{{P_t}{{\left( {L\left( {{d_{b,r}}} \right)} \right)}^2}}}{{{\sigma ^2}}},
	\end{align}
	where $ P_t $ is the transmit power of the BS, and $ {{d_{b,r}}} $ is the distance of the BS-IRS link.
	Moreover, to evaluate the performance of the channel estimation method, we define the normalized error of the estimated channel $ {{\bf{\hat G}}} $ as
	\begin{align}
	{\mathop{\rm NE}\nolimits} \big( {{\bf{\hat G}}} \big) = \mathop {\min }\limits_{{\bm{\delta }}:{\bm{\delta }}\left( n \right) =  \pm 1,\forall n} \frac{{{{\big\| {{\rm{diag}}\big( {\bm{\delta }} \big){\bf{\hat G}} - {\bf{G}}} \big\|}_F}}}{{{{\left\| {\bf{G}} \right\|}_F}}}.
	\end{align}
	
	\begin{figure}[htbp]
		\centerline{\includegraphics[width=0.38\textwidth]{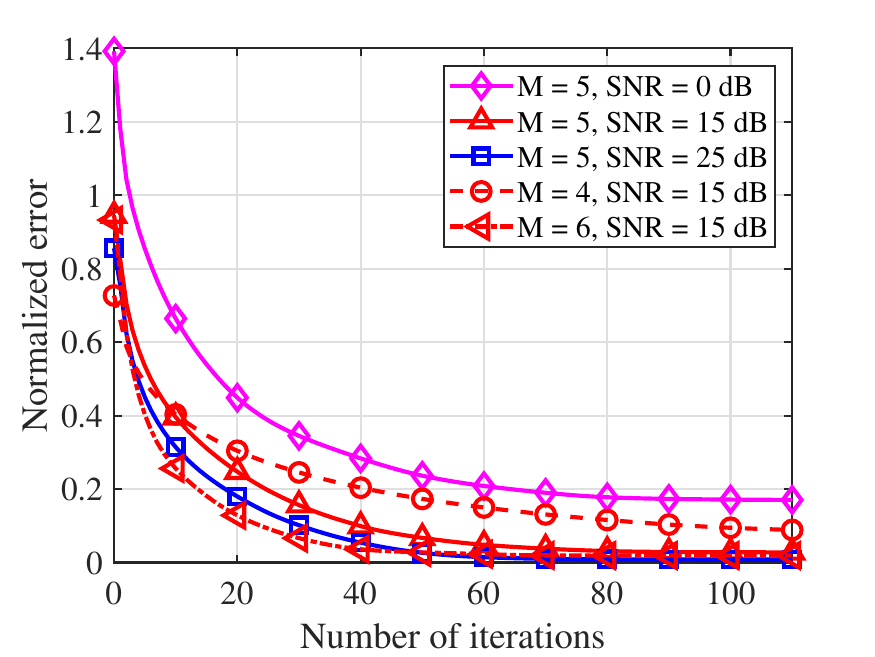}}
		\caption{The normalized error of the estimated channel $ {{\bf{\hat G}}} $ versus the number of iterations under different $ M $ and $ {{\mathop{\rm SNR}\nolimits} _r} $ settings. We set $ M_t = 1 $ and $ N = 25 $.}
		\label{fig_ChannelEsti_1}
	\end{figure}
	
	In Fig. \ref{fig_ChannelEsti_1}, we show the convergence behavior of the proposed separate BS-IRS channel estimation scheme under different $ M $ and $ {\rm{SN}}{{\rm{R}}_r} $ settings. In each iteration, the estimates of all the $ MN $ elements in $ {{\bf{\hat G}}} $ are refined from the first one to the last one. As shown in Fig. \ref{fig_ChannelEsti_1}, the normalized error of the estimated channel $ {{\bf{\hat G}}} $ drops as the coordinate descent-based algorithm proceeds. The proposed algorithm converges to a stable value after about $ 100 $ iterations for the worst case, which shows the difficulty of recovering $ {\left\{ {{g_{n,a}}} \right\}_{n \in {\cal N},a \in {\cal M}}} $ from the estimated value of $ {\left\{ {{g_{n,a}}{g_{n,b}}} \right\}_{n \in {\cal N},a \in {\cal M},b \in {\cal M}\backslash \left\{ a \right\}}} $. 
	From the red curves corresponding to $ {\rm{SN}}{{\rm{R}}_r} = 15{\text{ dB}} $ and $ M = 4,5 $ and $ 6 $, respectively, a larger number of BS antennas helps the algorithm converge faster. This is because a larger size of $ \bf{G} $ corresponds to more local optima of the objective function $ J $ in (\ref{ML_esti_2}), making it easier for the algorithm to converge to one of the locally optimal points. However, the convergence behavior of the proposed algorithm is less sensitive to the values of $ {\rm{SN}}{{\rm{R}}_r} $, showing that the number of iterations required for its convergence has no obvious monotonic relationship with the value of $ {\rm{SN}}{{\rm{R}}_r} $.
	
	\begin{figure}[htbp]
		\centerline{\includegraphics[width=0.38\textwidth]{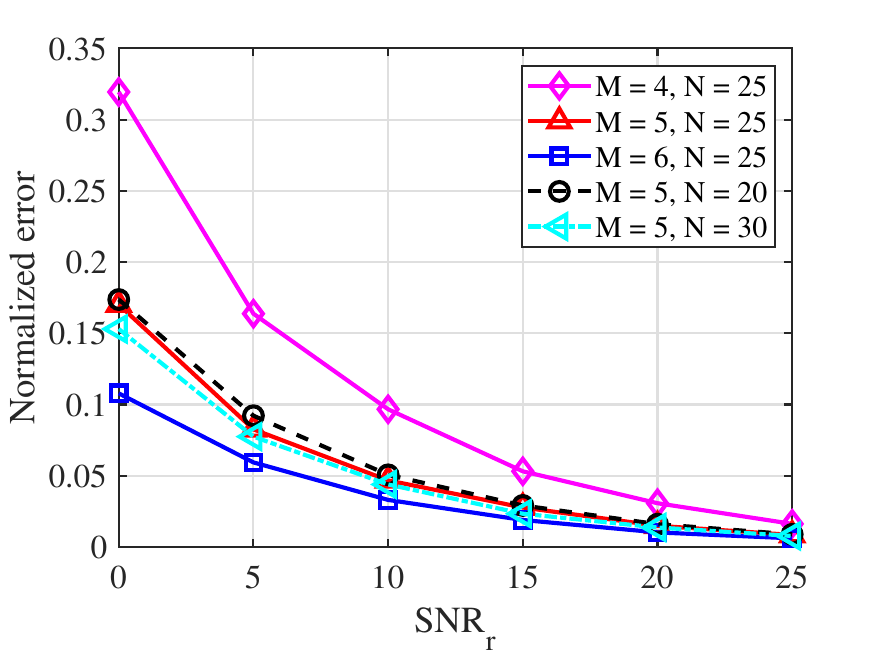}}
		\caption{The normalized error of the estimated channel $ {{\bf{\hat G}}} $ versus the received SNR under different $ M $ and $ N $ settings, where we set $ M_t = 1 $.}
		\label{fig_ChannelEsti_2}
	\end{figure}
	
	In Fig. \ref{fig_ChannelEsti_2}, we plot the normalized error of the estimated channel $ {{\bf{\hat G}}} $ versus the received SNR under different $ M $ and $ N $ settings. Obviously, the normalized error value decreases monotonically as $ {\rm{SN}}{{\rm{R}}_r} $ becomes large. In the case of the same number of IRS elements, the increase in the number of BS antennas will lead to more accurate channel estimation, which corresponds to the fact shown in Fig. \ref{fig_ChannelEsti_1} that the convergence is accelerated by a larger $ M $. In addition, we can also see that the performance increases with $ N $ for the same $ M $, although not very significantly. In general, a larger size of channel matrix $ \bf{G} $ yields a smaller estimation error.
	
	\begin{figure}[htbp]
		\centerline{\includegraphics[width=0.38\textwidth]{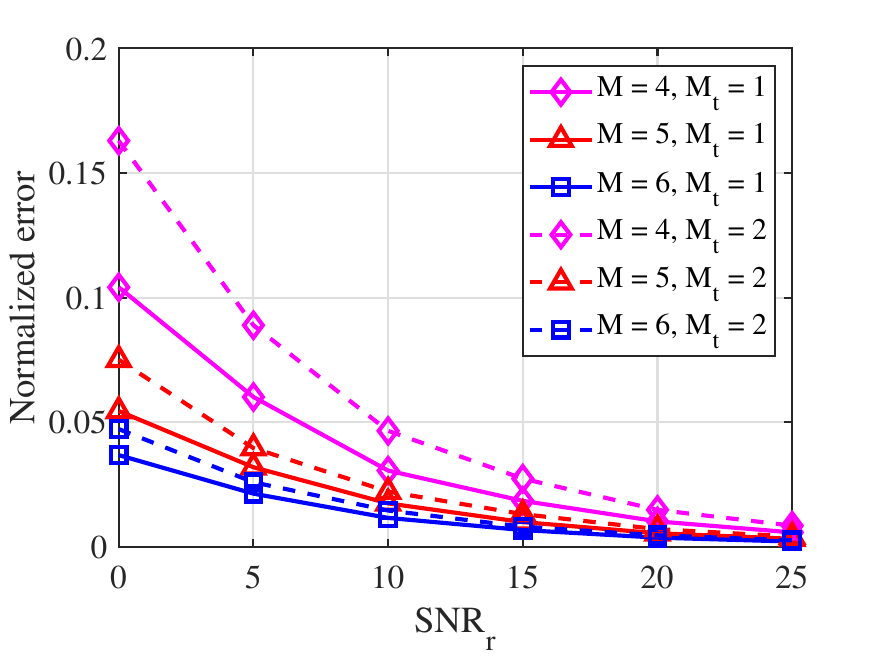}}
		\caption{Performance comparison of different transmit antenna numbers $ M_t $ for multiple $ M $ settings, where we set $ N = 25 $.}
		\label{fig_ChannelEsti_3}
	\end{figure}
	
	Fig. \ref{fig_ChannelEsti_3} compares the channel estimation performance of the cases when the number of transmit antennas are set to $ M_t = 1 $ and $ 2 $, respectively, for the same pilot overhead. Specifically, for $ M_t = 2 $ we set $ C = N{M_t} = 50 $, where $ C $ corresponds to the pilot overhead under each transmit and receive antenna combination, see the expression for $ {{\bf{\Phi }}^{\left( p \right)}} $ in (\ref{vec_y}). Moreover, considering that the number of transmit-receive antenna combinations is $ C\left( {M,2} \right) = \frac{{M\left( {M - 1} \right)}}{2} $, the total pilot overhead is calculated as $ PO = 25M\left( {M - 1} \right) $. To ensure that the pilot overhead is the same for both cases since the number of transmit-receive antenna combinations is $ C\left( {M,1} \right) = M $ when $ M_t = 1 $, we set $ C = \frac{{25M\left( {M - 1} \right)}}{M} = 25\left( {M - 1} \right) $ for this case.
	
	From Fig. \ref{fig_ChannelEsti_3}, it can be observed that for the same $ M $, the channel estimation accuracy corresponding to $ M_t = 1 $ is higher than that corresponding to $ M_t = 2 $, and the performance gap diminishes with an increase in $ M $. This observation suggests that in the absence of channel prior knowledge, i.e., multi-antenna beamforming cannot be achieved in this case, reducing the number of transmit antennas is a more cost-effective approach for channel estimation. In addition, by comparing the performance curves with the same $ M $, $ N $, and $ M_t $ in Fig. \ref{fig_ChannelEsti_3} and Fig. \ref{fig_ChannelEsti_2}, we can conclude that more pilot overhead helps to improve the channel estimation accuracy. This is due to the fact that the increase of pilot makes the obtained $ {{{\bm{\hat \omega }}}^{\left( p \right)}} $ in (\ref{omega_esti}) closer to the true value.

	\subsection{Target Localization}
	
	\begin{figure*}[t]
		\centerline{\includegraphics[width=0.9\textwidth]{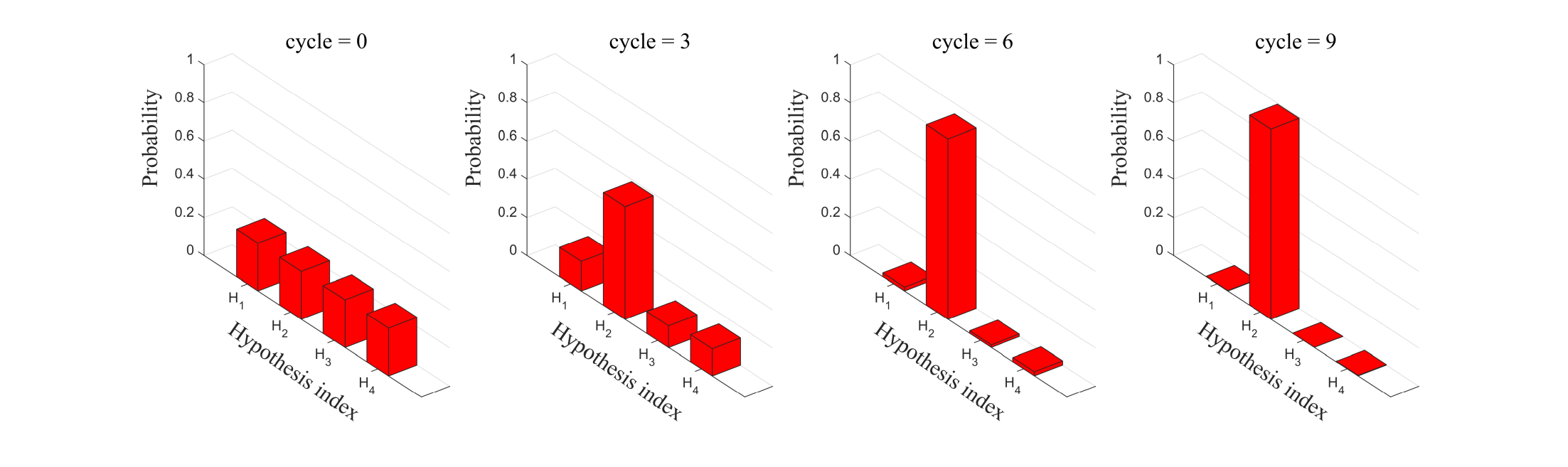}}
		\caption{Probability values of all the hypotheses in different cycles, where we set $ P_b = 50 \text{ W} $ and $ N_y = 4 $. The correct hypothesis is $ H_2 $.}
		\label{fig_Localization_1}
	\end{figure*}
	
	Fig. \ref{fig_Localization_1} illustrates the process of target localization by applying the hypotheses testing framework. The probabilities of all the hypotheses are initialized to be equal, while after several cycles, the probability of the correct hypothesis is obviously higher than those of the wrong hypotheses and finally approaches $ 1 $ at the $ 9 $-th cycle. The probability update trend for all hypotheses implies the effectiveness of our proposed target localization scheme. It is worth noting that because the BS-IRS channel is Rayleigh fading, the rate at which the probability of an incorrect hypothesis decays is not necessarily related to its distance from the correct hypothesis unless the incorrect hypothesis is within relative proximity of the correct hypothesis. Moreover, we can divide the angular range of interest into more grids to improve the localization accuracy.
	
	\begin{figure}[H]
		\centerline{\includegraphics[width=0.38\textwidth]{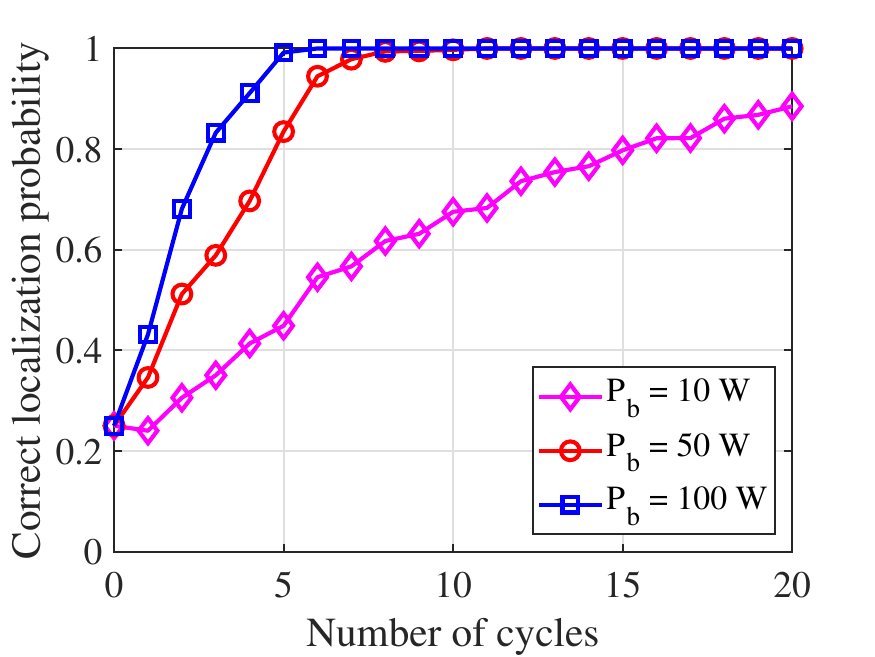}}
		\caption{The correct localization probability versus the number of cycles under different maximum transmit power $ P_b $ at the BS, where we set $ N_y = 4 $.}
		\label{fig_location_2}
	\end{figure}
	
	Fig. \ref{fig_location_2} presents the correct localization probability versus the number of cycles of our proposed localization scheme under different maximum power budget $ P_b $ at the BS.
	First, it can be observed that a larger $ P_b $ makes the correct localization probability approach 1 with fewer cycles required, and it corresponds to an elevated correct localization probability within the initial few cycles.
	Second, we observe that a larger $ P_b $ enhances the monotonicity of the performance curve, this is because the higher power localization waveform reduces the randomness impact caused by the received noise.
	Additionally, it is noted that for $ {P_b} = 10{\text{ W}} $, the correct localization probability obtained in the $ 1 $-th cycle becomes lower compared to the previous cycle. This performance degradation occurs because there is no joint optimization of the BS transmit waveform and IRS phase shifts in this particular localization cycle, thus leading to insufficient desired localization signal power compared to noise power, and consequently introducing randomness in localization performance.
	
	\begin{figure}[H]
		\centerline{\includegraphics[width=0.38\textwidth]{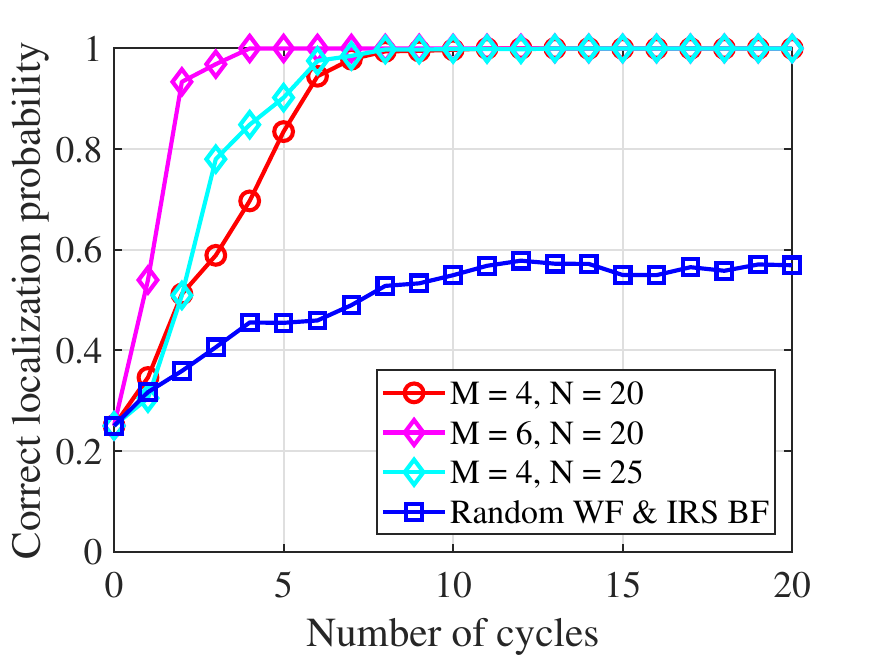}}
		\caption{The correct localization probability versus the number of cycles under different hardware configurations, where we set $ P_b = 50 \text{ W} $.}
		\label{fig_location_3}
	\end{figure}
	
	In Fig. \ref{fig_location_3}, we plot the correct localization probability versus the number of cycles under different hardware configurations and localization schemes. We can see that the localization performance can be enhanced by deploying a larger number of BS antennas and IRS elements. The reason behind this is twofold. First, the increased quantity of these components improves spatial resolution, thus better facilitating accurate target localization for the system. Moreover, augmenting both BS antennas and IRS elements increases the received useful echo signal power, while reducing the interference of the received noise signals. Besides, it is seen that our proposed target localization scheme is far superior to the ``Random WF \& IRS BF" benchmark scheme, where the BS transmit waveform and IRS beamforming are randomly selected and not optimized. The results demonstrate that the optimization of both BS transmit waveform and IRS phase shifts to increase the weighted sum ``predicted distance" between different hypotheses can effectively enhance the localization performance of the studied system.
	
	\section{Conclusion}
	In this paper, we have investigated the BS's wireless localization of a specific target in its NLoS region with the aid of an IRS. In particular, we have considered a special case where the BS-IRS CSI is unknown. The overall wireless localization procedure consists of two stages, namely, the separate BS-IRS channel estimation stage and the target localization stage.
	Specifically, in the BS-IRS channel estimation stage, we have adopted the simultaneous pilot transmission and reception scheme of the BS operating in the FDM and proposed an iterative coordinate descent-based algorithm to recover the BS-IRS
	channel. An incomplete BS-IRS channel matrix can be finally obtained due to the ``sign ambiguity issue". While in the target localization stage, based on the estimated incomplete BS-IRS channel matrix, we have employed the multiple hypotheses testing technique to obtain the target's location, and the complete BS-IRS channel matrix can also be conveniently determined. To further improve the target localization performance, we have optimized the BS transmit waveform and IRS phase shifts to maximize the weighted sum distance between hypotheses.
	Simulation results have verified the effectiveness of our proposed IRS-enabled target localization scheme. Moreover,
	we can also obtain the following two conclusions: 1) the overall target localization procedure can be used as the separate BS-IRS channel estimation scheme; 2) The target localization performance can be greatly improved by finely designing the BS transmit waveform and IRS phase shifts.

	\bibliographystyle{IEEEtran}
	\bibliography{mybibfile}

\end{document}